\documentclass{article}

\usepackage{PRIMEarxiv}

\usepackage[utf8]{inputenc} 
\usepackage[T1]{fontenc}    
\usepackage{url}            
\usepackage{booktabs}       
\usepackage{amsfonts}       
\usepackage{nicefrac}       
\usepackage{microtype}      
\usepackage{lipsum}
\usepackage{fancyhdr}       
\usepackage{graphicx}       
\graphicspath{{media/}}     

\usepackage{relsize,setspace}  
\usepackage{url}               
\usepackage{mathrsfs,amsmath} 
\usepackage{mathtools}  
\usepackage{amsfonts}  

\usepackage{algorithm}
\usepackage{algpseudocode} 

\usepackage[colorlinks=true,linkcolor=black, citecolor=blue, urlcolor=blue]{hyperref}

\usepackage[nameinlink,noabbrev]{cleveref} 

\usepackage{lineno} 
\usepackage{setspace}
\usepackage{indentfirst}
\usepackage{booktabs}

\usepackage{float}
\usepackage{graphicx}
\usepackage{caption}
\usepackage{subcaption}

\usepackage{tikz} 
\usepackage{diagbox} 

\usetikzlibrary{arrows,calc,fit}

\setkeys{Gin}{width=0.85\textwidth}

\newcommand{\BoldMath}[1]{\boldsymbol{\mathrm{ #1}}}

\newcommand{\Expect}{{\rm I\kern-.3em E}}

\usepackage{multicol}

\title{Information fusion and machine learning for sensitivity analysis using physics knowledge and experimental data
\thanks{\textit{\underline{Citation}}: 
\textbf{Kapusuzoglu, B., \& Mahadevan, S. Information fusion and machine learning for sensitivity analysis using physics knowledge and experimental data. \textit{Reliability Engineering \& System Safety} 214, 107712 (2021). DOI: \href{https://doi.org/10.1016/j.ress.2021.107712}{10.1016/j.ress.2021.107712}}
} 
}

\author{
  \textbf{Berkcan Kapusuzoglu}\thanks{Corresponding author: berkcan.kapusuzoglu@vanderbilt.edu} \quad and \quad \textbf{Sankaran Mahadevan} \\
  \vspace{0.2cm} \\
  Department of Civil and Environmental Engineering \\
  Vanderbilt University, Nashville, TN 37235, USA \\
  \texttt{berkcan.kapusuzoglu@vanderbilt.edu}, \quad \texttt{sankaran.mahadevan@vanderbilt.edu} \\
}

\begin{document}
\maketitle

\begin{abstract}
When computational models (either physics-based or data-driven) are used for the sensitivity analysis of engineering systems, the sensitivity estimate is affected by the accuracy and uncertainty of the model. This paper considers global sensitivity analysis (GSA) for situations where both a physics-based model and experimental observations are available, and investigates physics-informed machine learning strategies to effectively combine the two sources of information in order to maximize the accuracy of the sensitivity estimate. Two representative machine learning (ML) techniques are considered, namely, deep neural networks (DNN) and Gaussian process (GP) modeling, and two strategies for incorporating physics knowledge within these techniques are investigated, namely: (i) incorporating loss functions in the ML models to enforce physics constraints, and (ii) pre-training and updating the ML model using simulation and experimental data respectively. Four different models are built for each type (DNN and GP), and the uncertainties in these models are included in the Sobol’ indices computation. The DNN-based models, with many degrees of freedom in terms of model parameters and training options, are found to result in smaller bounds on the sensitivity estimates when compared to the GP-based models. The proposed methods are illustrated for additive manufacturing and lake temperature modeling examples.
\end{abstract}



\section{Introduction}\label{sec:Intro}
Computational models are often used to analyze the response of an engineering system for a variety of input realizations, since conducting experiments to directly measure the true response for many input realizations is often not affordable. However, the computational model is often an incomplete representation of the complex physical system, thus the system response prediction is affected by model uncertainty. In general, the uncertainty sources affecting system response prediction include (a) epistemic uncertainty due to lack of knowledge (arising from either data or model inadequacies), and (b) aleatory uncertainty due to the inherent variability in the system properties or the external inputs. Global sensitivity analysis (GSA)~\cite{saltelli2008global} aims to provide a quantitative assessment of the relative contribution of each uncertainty source to the uncertainty in the model response~\cite{saltelli1999quantitative,mara2012variance,borgonovo2016sensitivity}.

Much of the GSA literature has focused on variability in the inputs and their effects on output variability; the extension of GSA to include epistemic uncertainty sources (data, model) is recent and sparse~\cite{sankararaman2013separating,li2016relative, le2014bayesian,oakley2004probabilistic,marrel2009calculations}. Model outputs can have uncertainty even for a fixed input when there exists model uncertainty. When the model is computationally expensive, it is often replaced with a surrogate model to facilitate the estimation of Sobol’ indices, since such computation requires many input-output samples from the model; the surrogate model introduces additional uncertainty. Several types of surrogate models are used in the literature, e.g., polynomial chaos expansion (PCE), Gaussian process (GP) regression, neural networks, etc., to train a parametric relationship between the inputs and the outputs. The quality and quantity of the training data affect the accuracy of these surrogate models, which directly affects the uncertainty in the model output~\cite{le2014bayesian,oakley2004probabilistic,marrel2009calculations}. Thus, it is important to also include the contribution of surrogate model uncertainty to the output uncertainty in GSA. In Le Gratiet et al~\cite{le2014bayesian} for example, the Gaussian process surrogate model uncertainty is included in the Sobol’ index estimates using multiple realizations of the GP model prediction, which helps to construct prediction intervals for the Sobol’ index estimates.

Expanding GSA to consider both aleatory and epistemic uncertainty sources is beneficial in supporting resource allocation decisions. If the contribution of epistemic uncertainty is found to be significant, then it may be valuable to collect more data or refine the physics model to reduce the epistemic uncertainty and thus its contribution to the output uncertainty. Several GSA studies have developed auxiliary variable-based approaches to include both aleatory and epistemic uncertainty sources at a single level instead of using nested simulations, thus achieving both computational efficiency and direct ranking of the different sources of uncertainty to support resource allocation decision-making. The auxiliary variable is used to transform one-to-many input-output mapping to one-to-one mapping, thus facilitating the computation of Sobol’ indices for both aleatory and epistemic sources~\cite{sankararaman2013separating}. This idea is expanded in~\cite{li2016relative} to include several epistemic sources, such as input statistical uncertainty, surrogate model error, physics model discrepancy, and numerical solution error, and to systems with time series inputs and outputs.

Three scenarios of model and data availability can be considered for GSA: (1) use of a physics-based computational model alone, (2) use of available input-output data alone (either from experiments or previous simulations), or (3) use of both physics model and available experimental data. A straightforward model-based approach to estimate Sobol’ indices is to use a double-loop Monte Carlo simulation (MCS) ~\cite{sobol2001global}. In order to reduce the cost associated with the double-loop MCS, analytical, spectral and efficient sampling-based methods have been developed. The methodology developed by Sudret~\cite{sudret2008global} approximates the original physics model by a PCE and estimates the Sobol’ indices by using the PCE coefficients. Chen et al.~\cite{chen2005analytical} proposed analytical formulas to compute Sobol’ indices using a GP surrogate model with input variables that follow normal or uniform distributions. The improved FAST method~\cite{tarantola2006random} combines the classical FAST method~\cite{saltelli1999quantitative} with random balanced design~\cite{satterthwaite1959random} for generating samples to evaluate Sobol’ indices.

In some problems, input-output data may be already available instead of having to simulate a physics model expressly for the purpose of GSA. Such data may be available from experiments, or real-world observations, or Markov Chain Monte Carlo (MCMC) sampling during Bayesian model calibration, or MC sampling during reliability analysis, etc. In such cases, data-driven methods have been proposed to directly compute the Sobol’ indices based on available input-output samples instead of simulation runs of the physics model. A GSA method based on ANOVA using factorial design of experiments is developed by Ginot et al.~\cite{ginot2006combined}. The proposed method results in same values as the Sobol’ index since the variance decomposition used in the Sobol’ index estimations is same as the one used in the classical ANOVA~\cite{archer1997sensitivity}. In high dimensional problems, even the use of a surrogate model for GSA, which repeatedly executes the code by suppressing some variables and running through the range of other variables, may be computationally demanding since the number of executions of the code increases rapidly with the number of inputs~\cite{le2014bayesian,hu2019probability,marrel2008efficient,o2006bayesian}. The computational cost of most sample-based methods is proportional to the number of model inputs. Li and Mahadevan~\cite{li2016efficient} proposed a modularized method, which has a computational cost that is not proportional to the model input dimension, to estimate the first-order Sobol’ indices based on stratification of available input-output samples. DeCarlo et al.~\cite{decarlo2018efficient} proposed an importance sampling approach by introducing weights to different data points to estimate Sobol’ indices from available data using Sobol’ sequences to reduce the number of simulations; this method computes both first-order and higher order indices, and is able to include correlated inputs. Approximations to the joint probability distribution of inputs and outputs such as multivariate Gaussian, Gaussian copula, and Gaussian mixture have recently been found to give rapid estimation of Sobol’ indices ~\cite{hu2019probability}.

The third scenario is of interest in this paper, where both a physics-based model and some experimental or real-world data are available. One option, if adequate data is available, is to simply build a regression or machine learning (ML) model based on the observation data, and use this model to perform GSA. Multiple recent studies have pursued data-driven ML models in situations where abundant experimental data or real-world observations are available due to advances in modern sensing techniques. Generally, the construction of data-driven ML models does not require in-depth knowledge of the complex physics inherent in the physical process. ML models can learn complex systems using available observations, but the accuracy of these models depends on the quality and quantity of the data. If the available data is sparse, then the complexity of the process may not be fully captured. Further, since purely data-driven ML models do not explicitly consider physical laws, they can produce physically inconsistent results. In such cases, incorporating physics knowledge within ML models may improve the accuracy and efficiency of GSA computations. The combined use of physics-based and ML models has been shown to achieve more accurate and physically consistent predictions by leveraging the advantages of each method~\cite{karpatne2017physics,jia2020physics,willard2020integrating,berkcan2020piml}.
 
In this work, we incorporate physics knowledge into the ML models to better capture the physics of the process by leveraging physical laws while improving the generalization performance of data-driven models. Two types of strategies are considered for incorporating physics knowledge within ML models: (1) incorporating loss functions in the ML model training to enforce physics constraints, and (2) pre-training the ML model with data generated by the physics model and then updating it with experimental data. Note that the first strategy does not use the physics model but only constraints for the output to obey physical requirements; whereas, the second strategy explicitly uses the physics computational model. Two types of ML models are considered in this paper, namely, Gaussian process (GP) and deep neural network (DNN). These two models are selected in order to represent two different kinds of available ML techniques; the GP model is one of the surrogate models commonly used in uncertainty quantification (UQ) studies, and DNN belongs to the emerging class of deep learning algorithms revolutionizing the field of artificial intelligence, spurred by recent advances in sensing, communication and computational resources. Four different physics-informed machine learning (PIML) models are developed for each type (i.e., GP or DNN) to predict the output quantity of interest (QoI), through combinations of the two strategies. The resulting GSA procedure incorporates the effect of uncertainty in the ML or PIML model, and the various models and strategies are compared in terms of accuracy and uncertainty in the GSA results and their computational demand.

In summary, the contributions of this paper are as follows:
\begin{itemize}
    \item Physics knowledge and experimental observations are fused in order to maximize the accuracy of sensitivity estimates.
    \item Two PIML strategies and their combinations are investigated for global sensitivity analysis using both physics knowledge and experimental data.
    \item Four different models are built for each of GP and DNN, and the uncertainties in these models are included in the Sobol’ indices computation.
    \item The accuracy, uncertainty and computational effort of different options for the ML and PIML models are evaluated and compared.
\end{itemize}

The outline of the rest of the paper is as follows. Section~\ref{Sec:Background} provides background information on related methods. Section~\ref{Sec:Methods} presents the proposed methodology. Two numerical examples are presented in Section~\ref{Sec:Results} to illustrate the proposed methodology and draw insights on the performance of various PIML strategies and models. Concluding remarks are provided in Section~\ref{Sec:Conclusion}.

\section{Background}\label{Sec:Background}
This section introduces each of the basic techniques used in developing the proposed methodology, namely variance-based GSA, Gaussian process (GP) surrogate modeling, and deep neural networks (DNN). These techniques are well established with extensive literature, therefore only a brief introduction is given here. 

\subsection{Variance-based GSA}\label{Sec:Sensitivity}
Consider a deterministic real integrable one-to-one system response function $\mathrm{Y}=f(\BoldMath{X})$, where $f(\boldsymbol{\cdot})$ is the computational model, $\BoldMath{X}=\{\mathrm{X}_1, ..., \mathrm{X}_k\}$ are mutually independent model inputs, and $\mathrm{Y}$ is the model output. As shown in \cite{sobol2001global}, the variance of $\mathrm{Y}$ can be decomposed as
\begin{align}
    V(\mathrm{Y}) = & \sum_i^k V_i + \sum_{i_1}^k\sum_{i_2=i_1+1}^k V_{i_1i_2} + \sum_{i_1}^k\sum_{i_2=i_1+1}^k\sum_{i_3=i_2+1}^k V_{i_1i_2i_3} + ... + V_{12...k}
    \label{eq:SobolVar}
\end{align}
where $V_i$ is the variance of $\mathrm{Y}$ due to $\mathrm{X}_i$ alone, and $V_{i_1...i_p} (p\geq2)$ indicates the variance of $\mathrm{Y}$ caused by the interaction of $\{\mathrm{X}_{i_1},...,\mathrm{X}_{i_p}\}$.

The Sobol' indices are defined by dividing both sides of Eq.~\eqref{eq:SobolVar} with $V(\mathrm{Y})$
\begin{align}
    1= & \sum_i^k S_i + \sum_{i_1}^k\sum_{i_2=i_1+1}^k S_{i_1i_2} + \sum_{i_1}^k\sum_{i_2=i_1+1}^k\sum_{i_3=i_2+1}^k S_{i_1i_2i_3} + ... + S_{12...k}
    \label{eq:SobolIndex}
\end{align}
where $S_i$ is the first-order or main effects index that assesses the contribution of $\mathrm{X}_i$ individually to the variance of the output $\mathrm{Y}$ without considering interactions with other inputs. The higher-order indices $S_{i_1...i_p} (p\geq2)$ in Eq.~\eqref{eq:SobolIndex} measure the contributions of the interactions of $\{\mathrm{X}_{i_1},...,\mathrm{X}_{i_p}\}$.

The first-order index $S_i$ is defined as follows:
\begin{equation}
    S_i = \frac{V_i}{V(\mathrm{Y})} = \frac{V_{X_i}(E_{\BoldMath{X}_{-i}}(\mathrm{Y}|\mathrm{X}_i))}{V(\mathrm{Y})}
    \label{eq:FirstIndex}
\end{equation}
where $\BoldMath{X}_{-i}$ are all the model inputs other than $\mathrm{X}_i$.

The overall contribution of $\mathrm{X}_i$ considering an individual input and its interactions with all other inputs is measured by the total effects index $S_i^T$:
\begin{equation}
    S_i^T = 1 - \frac{V_{-i}}{V(\mathrm{Y})} = \frac{V_{\BoldMath{X}_{-i}}(E_{\mathrm{x}_i}(\mathrm{Y}|\BoldMath{X}_{-i}))}{V(\mathrm{Y})}.
    \label{eq:TotalIndex}
\end{equation}

The computation of $S_i$ analytically is nontrivial since $E_{\BoldMath{X}_{-i}}(\boldsymbol{\cdot})$ requires multi-dimensional integrals. A basic sampling-based approach is to use double-loop sampling~\cite{sobol2001global}. Several approaches to reduce the computational cost were mentioned in Section~\ref{sec:Intro}. One of these approaches of particular relevance to this paper is to replace the original computational model $f(\boldsymbol{\cdot})$ by a surrogate model and use this surrogate model in GSA~\cite{gratiet2016metamodel,marrel2012global,xiu2002wiener,janon2014uncertainties, hu2015mixed}. This approach will be addressed further in Section~\ref{Sec:Methods}.

\subsection{Gaussian process surrogate modeling}\label{Sec:GP}
A Gaussian process (GP) surrogate model (or kriging) approximates a response function $y = G(\mathbf{x})$ over the domain of input $\mathbf{x}$ as a Gaussian random process with a mean function $m(\mathbf{x})$ and a covariance function $k(\mathbf{x},\mathbf{x^{\prime}})$, which describes the deviation of the model from the trend
\begin{equation}
    \mathbf{y}(\mathbf{x}) \sim \mathcal{GP}\big(m(\mathbf{x}),  k(\mathbf{x},\mathbf{x^{\prime}})\big).
    \label{eq:GP}
\end{equation}

Given a set of training data \{$\mathbf{X}_T,\mathbf{Y}_T$\} and the input $\mathbf{X}_P$, where prediction is desired, the conditional probability distribution of the output $\mathbf{Y}_P$ follows a multivariate Gaussian distribution~\cite{williams2006gaussian} as
\begin{align}\label{eq:predictiveDist}
    & \mathbf{Y}_P|\mathbf{X}_P,\mathbf{X}_T,\mathbf{Y}_T \sim \mathcal{N}(\boldsymbol{\mu}, \Sigma) \\ \notag
    & \boldsymbol{\mu}=m(\mathbf{X}_P)+\Sigma_{PT}(\Sigma_{TT}+\sigma_{obs}^2\boldsymbol{I} )^{-1}\big(\mathbf{Y}_T- m(\mathbf{X}_T)\big)\\ \notag 
    & \Sigma=\Sigma_{PP} - \Sigma_{PT}(\Sigma_{TT}+\sigma_{obs}^2 \boldsymbol{I})^{-1}\Sigma_{PT}^T
\end{align}
where $\boldsymbol{\mu}$ is the mean vector of the prediction $\mathbf{Y}_P$ conditioned on the training data, and $\Sigma$ is the conditional covariance matrix of $\mathbf{Y}_P$; $\Sigma_{PT}$ is the covariance matrix between the prediction data $\mathbf{X_P}$ and each of the training points \{$\mathbf{X_T} = \boldmath{x_1}, \boldmath{x_2}, ..., \boldmath{x_{n}}$\}; $ \Sigma_{TT}$ is the $n \times n$ covariance matrix of the training data; $\Sigma_{PP}$ is the unconditional covariance matrix of $\mathbf{Y}_P$; $\sigma_{obs}^2$ is the variance of observation/measurement error (also called noise variance), and $\boldsymbol{I}$ is the identity matrix.


The mean function $m(\cdot)$ is often formulated as a polynomial function of inputs. The covariance function $k(\cdot)$ can be formulated by using a covariance function based on the desired properties (order of continuity, stationary/non-stationary, isotropic/anisotropic). A squared exponential correlation function with separate length scale parameters $l_i$ for each input dimension has often been used in the literature:
\begin{equation}
k(\mathbf{x}, \mathbf{x^{\prime}})=\sigma_f^2\exp \left[-\sum_{i=1}^{\rm M} \frac{\left(x_{i}-x^{\prime}_{i}\right)^{2}}{2l_i}\right] + \sigma_{obs}^2\delta(\mathbf{x},\mathbf{x}^{\prime})
\end{equation}
where $\sigma_f^2$ is the signal/process variance and defines the maximum allowable covariance, and $\delta(\mathbf{x},\mathbf{x}^{\prime})$ is the Kronecker delta function.


The hyperparameters of the GP model considering a zero mean function, i.e., $\mathbf{\Theta}=\{l,\sigma_f,\sigma_{obs}\}$, are inferred from the training data. A common method is to maximize the log marginal likelihood function, which is defined as
\begin{align}
\mathrm{log}\ p(\mathbf{Y}_T | \mathbf{X}_T; \mathbf{\Theta}) = -\frac{1}{2}\mathbf{Y}_T(\Sigma_{TT}+\sigma_{obs}^2\boldsymbol{I})^{-1}\mathbf{Y}_T - \frac{1}{2}\mathrm{log}|\Sigma_{TT}+\sigma_{obs}^2\boldsymbol{I}|-\frac{n}{2}\mathrm{log}2\pi.
\label{eq:loglik}
\end{align}

\subsection{Deep neural networks}\label{sec:DNN}
\begin{figure}[!ht]
	\centering
	\includegraphics[width=0.5\textwidth]{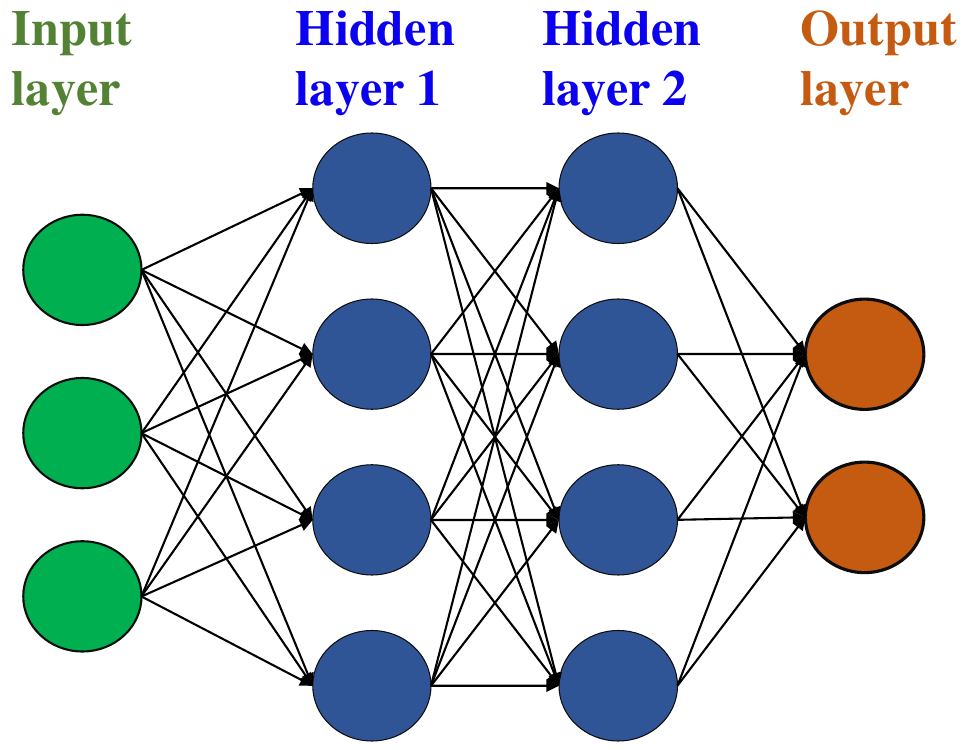}
	\caption{A deep neural network with two hidden layers.}
	\label{fig:DNN}
\end{figure}

In recent years, due to the confluence of advanced sensing and imaging techniques, big data processing techniques, enormous computational power and the internet, rapid advances are being made in developing sophisticated data-driven machine learning models, particularly neural networks. A deep neural network (DNN) is composed of multiple hidden layers and has four major components: neuron, activation function, cost function, and optimization. Figure~\ref{fig:DNN} shows a neural network consisting of three inputs, two hidden layers, each having four neurons, and two output neurons. The values of various input variables of a particular neuron are multiplied by their associated weights, then the sum of the products of the neuron weights and the inputs are calculated at each neuron. The summed value is passed through an activation function that maps the summed value to a fixed range before passing these signals on to the next layer of neurons.

The predictions of the DNN after forward propagation, $\mathbf{\hat{Y}}$, are compared against the observations, $\mathbf{Y}_{obs}$, by defining a loss function (e.g., root mean squared error (RMSE); $\mathcal{L}_{\rm RMSE}(\mathbf{Y}_{obs}, \mathbf{\hat{Y}})=\sqrt{\sum_{i=1}^{n} (\mathrm{y}_{obs,i}- \mathrm{\hat{y}}_{i})^2/n}$), which measures how far off the predictions are from the observations for the $n$ training samples. Backpropagation algorithms are employed to keep track of small perturbations to the weights that affect the error in the output and to distribute this error back through the network layers by computing gradients for each layer using the chain rule. In order to minimize the value of the loss function, necessary adjustments are applied at each iteration to the neuron weights in each layer of the network. These procedures are performed at each iteration until the loss function converges to a stable value.

\section{Proposed methodology}\label{Sec:Methods}
The proposed methodology for sensitivity analysis, using both physics knowledge and experimental data, consists of the following steps:
\begin{enumerate}
    \item Identification of PIML strategies
    \item Implementation of PIML strategies in ML models
    \item Variance quantification in ML model prediction
    \item Sobol' indices computation with ML model prediction variance
\end{enumerate}

The following subsections describe these steps in detail.

\subsection{Identification of PIML strategies}\label{sec:PIML}
PIML models seek to incorporate physics knowledge or constraints within the data-driven ML models. When a mechanistic, physics-based model is also available, complementary strengths of both mechanistic and ML models can be leveraged in a synergistic manner~\cite{willard2020integrating}. In the latter case, the aim is to improve the predictions beyond that of physics-based models or ML models alone by coupling physics-based models with ML models. Thus two different strategies to combine physics knowledge and ML models can be considered: (1) incorporate physics constraints in the ML models, and (2) pre-train and update the ML models using physics model input-output and experimental data, respectively.

\subsubsection{Strategy 1: Enforcing physics constraints}
A direct strategy to enforce physics constraints in ML model predictions is by including the constraints within the loss function used in training the ML model~\cite{karpatne2017physics}. Thus, while training a PIML model with inputs $\mathbf{X}_{obs}$ and outputs $\mathbf{Y}_{obs}$, the physical constraints can be incorporated as additional penalty terms in the loss function:
\begin{align}\label{eq:lossfunc}
    \mathcal{L} = \mathcal{L}_{\rm ML} + \lambda_{\rm phy}\mathcal{L}_{\rm phy}(\mathbf{\hat{Y}}),
\end{align}
where $\mathcal{L}_{\rm ML}$ is the log marginal likelihood of the data for a GP model:
\begin{equation}
    \mathcal{L}_{\rm GP} = \mathrm{log}\ p(\mathbf{Y}_{obs} | \mathbf{X}_{obs}; \mathbf{\Theta})
\end{equation}
with experimental observations \{$\mathbf{X}_{obs},\mathbf{Y}_{obs}$\} being the training data for the GP model. For a DNN $\mathcal{L}_{\rm ML}$ represents training a loss function that evaluates a supervised error, e.g., root mean squared error (RMSE):
\begin{equation}\label{eq:lossfnc_dnn}
\mathcal{L}_{\rm DNN}(\mathbf{Y}_{obs}, \mathbf{\hat{Y}})=\sqrt{\sum_{i=1}^{n} \frac{(Y_{obs,i}- \hat{Y}_{i})^2}{n}}
\end{equation}
which measures the accuracy of predictions $\mathbf{\hat{Y}}$ for $n$ training samples. (Note that for the GP model, the likelihood is maximized, whereas for the DNN model, the RMSE is minimized). The additional physics constraint loss function $\mathcal{L}_{\rm phy}$ in the second term of Eq.~\eqref{eq:lossfunc} is weighted by a hyperparameter $\lambda_{\rm phy}$; the value of $\lambda_{\rm phy}$ controls the strength of the physics constraint enforcement. The inclusion of the second term ensures physically consistent model predictions and helps to reduce the generalization error, which is a measure of how accurately a model is able to predict the output QoI for previously unseen data~\cite{karpatne2017physics}.

The physical inconsistencies in the model predictions are evaluated using the physics constraint loss term. The generic forms of these physical relationships can be expressed using the following constraints:
\begin{align}\label{eq:lossfnc_phy}
    \mathcal{F}_{1}(\mathbf{\hat{Y}},\mathbf{\Gamma}) = 0, \\ \notag
    \mathcal{F}_{2}(\mathbf{\hat{Y}},\mathbf{\Gamma}) \leq 0.
\end{align}
where $\mathbf{\Gamma}$ denotes other variables or thresholds that define the physics constraint regarding the model output $\mathbf{\hat{Y}}$. Equation.~\ref{eq:lossfnc_phy} indicates that the constraints may take the form of equalities or inequalities. These equations can involve algebraic relationships or partial differentials of $\mathbf{\hat{Y}}$ and/or $\mathbf{\Gamma}$. The physics-based loss functions for these equations can be defined as:
\begin{align}
    & {}\mathcal{L}_{{\rm phy}}(\mathbf{\hat{Y}}) = ||\mathcal{F}_{1}(\mathbf{\hat{Y}},\mathbf{\Gamma})|| + \rm{ReLU}(\mathcal{F}_{2}(\mathbf{\hat{Y}},\mathbf{\Gamma})),
\end{align}
where $\rm{ReLU}(x)=max(0,x)$ represents the rectified linear unit function and it acquires value when a threshold is violated in the inequality constraint, i.e., it penalizes the optimization when $\mathcal{F}_{2} > 0$. It can also be used to penalize deviations from a desired physically consistent relationship among multiple outputs $\mathbf{\hat{Y}}$~\cite{muralidhar2018incorporating}.

\subsubsection{Strategy 2: Pre-training and Updating}\label{sec:Pretrain}
The ML model output accuracy and uncertainty are dependent on the quality and quantity of the available training data. In some systems, the high cost associated with conducting experiments makes it infeasible to have adequate amount of training data to build purely data-driven models. Thus, it may be desirable to combine the physics-based model and available experimental data in seeking to maximize the accuracy of the sensitivity estimates. When the experiments are expensive, they can only be conducted for a few values of the inputs, whereas it might be possible to run the physics-based model for a larger set of input values. In that case, the simulation data can be used to pre-train an ML model, which is used as the initial model to be updated with experimental observations. Further, training of ML models requires the choice of initial values of the model parameters. The transfer of physical knowledge using a pre-trained ML model can prevent poor initialization due to lack of knowledge regarding the initial choice of ML model parameters prior to training.

Since the pre-training based on the physics-based model can use a large amount of training data (with multiple input parameter combinations) over a wide range of values, the pre-training may also help the eventual ML model to have wider generalization beyond experimental data. In the numerical example in Section~\ref{Sec:Results}, the pre-training strategy exercises the physics model over 1310 input combinations, whereas only 39 experiments are available. However, if the physics model is computationally expensive, then the advantage of the pre-training strategy in using a larger input data set (for physics model runs) compared to the experiments becomes limited.

\begin{figure}[!ht]
	\centering
	\includegraphics[width=0.7\textwidth]{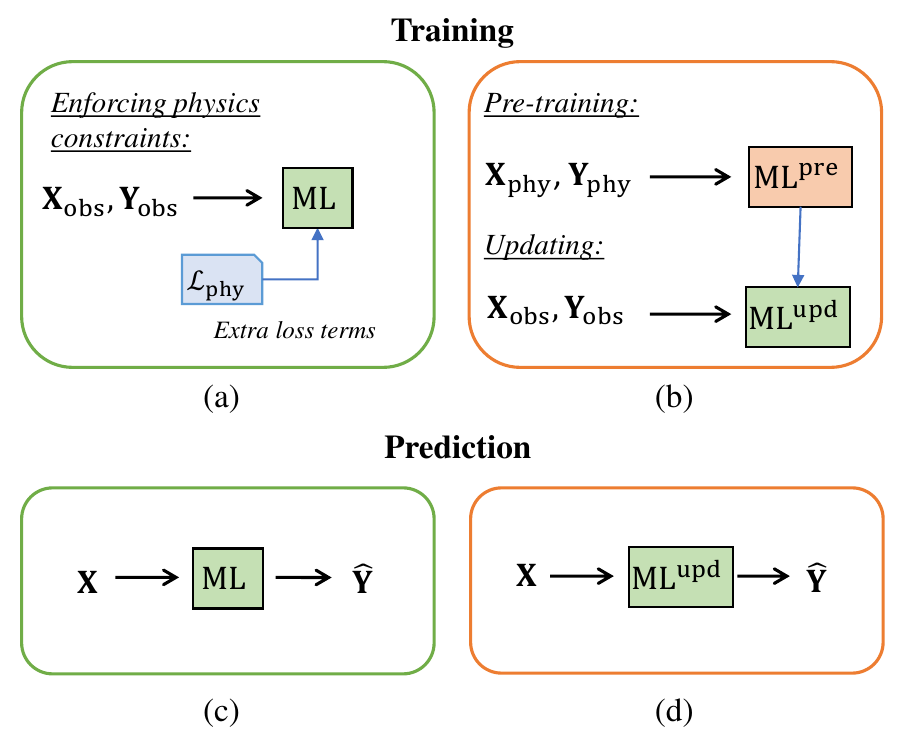}
	\caption{PIML strategies: (a) incorporating physics-based loss functions in the ML models to enforce physics constraints, (b) pre-training an ML model with physics model input-output ($\mathbf{X_{\mathrm{phy}}}$, $\mathbf{Y_{\mathrm{phy}}}$) and updating it with experimental training data ($\mathbf{X_{\mathrm{obs}}}$, $\mathbf{Y_{\mathrm{obs}}}$), (c) the trained ML model prediction $(\mathbf{\hat{Y}})$, (d) updated ML model prediction $(\mathbf{\hat{Y}})$.}
	\label{fig:diagram}
\end{figure}

The two proposed strategies to predict the QoI are shown in Fig.~\ref{fig:diagram}. Figure~\ref{fig:diagram}(a) shows the first method, where the physical knowledge is included through constraints within the loss function of an ML trained with only experimental data. Figure~\ref{fig:diagram}(b) shows the second method, where an ML model is first trained with data generated using the physics-based model and then updated using experimental data. Figures~\ref{fig:diagram}(c) and~\ref{fig:diagram}(d) show the trained ML model predictions ($\boldsymbol{\hat{Y}}$) for the two proposed strategies, respectively. The proposed PIML strategies can be applied to any physical system by leveraging the related physical constraints or physics-based models.

\subsection{Implementation of PIML strategies in ML models}\label{Sec:GP_DNN_PIML}
Based on the proposed two strategies to incorporate physics knowledge into the ML model, four separate ML models can be constructed for each type of surrogate model considered here (i.e., GP and DNN):
\begin{multicols}{2}
    \begin{enumerate}
        \item $\mathbf{\rm GP}$
        \item $\mathbf{\rm GP^{\mathcal{L}_{\rm phy}}}$
        \item $\mathbf{\rm GP^{\rm upd}}$
        \item $\mathbf{\rm GP^{\rm upd_1, \mathcal{L}_{\rm phy}}}$
        \item $\mathbf{\rm DNN}$
        \item $\mathbf{\rm DNN^{\mathcal{L}_{\rm phy}}}$
        \item $\mathbf{\rm DNN^{\rm upd}}$
        \item $\mathbf{\rm DNN^{\rm upd, \mathcal{L}_{\rm phy}}}$
    \end{enumerate}
\end{multicols}

These different models cover the following options: model trained with experimental data alone, models trained with PIML strategies 1 or 2 alone, and models trained with both PIML strategies together. The implementations of PIML strategies 1, 2, and their combination are different for the GP models vs. the DNN models. The following subsections describe how the PIML strategies can be implemented for each of the above models. 

\subsubsection{Implementation of PIML in GP models}\label{Sec:GP_PIML}
In Model 1, denoted as $\mathbf{\rm GP}$, only experimental observations are used for training. The hyperparameters of the GP model (process variance, correlation length scale along each input dimension, and trend function coefficients, and also measurement error variance if unknown) are optimized during training by maximizing the log marginal likelihood function shown in Eq.~\eqref{eq:loglik}. In calculating the likelihood, the difference between the true response of the system $\mathbf{Y}_{\rm true}$ and the observed response $\mathbf{Y_{\mathrm{obs}}}$ is attributed to the observation error $\epsilon_{\rm obs}$, which is often treated as a zero-mean Gaussian random variable with variance $\sigma_{\rm obs}^2$.

Model 2, denoted as $\mathbf{\rm GP^{\mathcal{L}_{\rm phy}}}$, incorporates the first PIML strategy by enforcing physics constraints during the optimization of the GP model hyperparameters. More specifically, the physics constraints are included during the maximization of the log marginal likelihood function (Eq.~\eqref{eq:loglik}) while inferring the hyperparameters of the GP model. Thus, the training of Model 2 is achieved by maximizing the function in Eq.~\eqref{eq:lossfunc_gp}:
\begin{align}\label{eq:lossfunc_gp}
    \mathcal{L}_{\rm GP} = \mathrm{log}\ p(\mathbf{Y}_{obs} | \mathbf{X}_{obs}; \mathbf{\Theta}) - \lambda_{\rm phy}\mathcal{L}_{\rm phy}(\mathbf{\hat{Y}}),
\end{align}
where $\mathbf{\hat{Y}}$ is the GP model prediction. Note that since $\mathcal{L}_{\rm GP}$ is to be maximized, the second term corresponding to the physics constraint has a negative sign. Gaussian process modeling under constraints has been studied in the literature~\cite{COUSIN2016631,AFST_2012_6_21_3_529_0,golchi2015monotone,lopez2018finite,riihimaki10a}. Veiga et al. ~\cite{AFST_2012_6_21_3_529_0} developed a framework that incorporates bound, monotonicity and convexity constraints in GP modeling. Golchi et al.~\cite{golchi2015monotone} developed a Bayesian approach to GP modeling that incorporates the monotonicity constraint. The need to obtain the monotonicity information at each of the points in the derivative input set can slow down the computation as the input dimension increases since the size of the covariance matrix depends on the input dimension. This paper does not use the methods described above. Instead, the paper proposes a different method that penalizes violations of the physics constraints by introducing a regularization term in the likelihood function. To the best of our knowledge, this is the first study to apply the penalty approach to the likelihood function of the GP model. Further, the computational effort of the proposed method (i.e., the calculation of the regularization term) does not increase with the problem size since the regularization term does not need the inverse of the covariance matrix.

The current work considers two different approaches for the second PIML strategy. Both approaches use a pre-trained model, obtained using the physics model input-output data. In the first approach, the model parameters are updated using the experimental data; and in the second approach, a discrepancy correction term is added to the pre-trained model. The first step of pre-training using data generated by the physics model can be thought of as similar to a lower fidelity model, and the second step of improving the pre-trained model (either by parameter updating or by adding a discrepancy term) can be thought of as similar to incorporating higher fidelity data (experimental data, in this case) to improve the model. Various multi-fidelity modeling strategies with different combinations of the low- and high-fidelity models have been studied in the literature, such as filtering, fusion, and adaptation~\cite{peherstorfer2018survey}. The discrepancy correction approach pursued here adopts the simplest strategy, namely, additive correction, where a model discrepancy term is added to the low-fidelity model~\cite{absi2016multi}:
\begin{equation}
    f_{HF}(\mathbf{X}) = f_{LF}(\mathbf{X}) + \delta(\mathbf{X};\boldsymbol{\theta_{\delta}})
\end{equation}
where $f_{HF}(\cdot)$ and $f_{LF}(\cdot)$ are the high and low fidelity models, respectively, $\mathbf{X}$ is the input to the model, and $\theta_{\delta}$ are the parameters of the discrepancy correction term. The correction term can be obtained using any suitable surrogate modeling technique.

Model 3, denoted as $\mathbf{\rm GP^{\rm upd}}$, pursues the first approach of the second PIML strategy, where a GP model is pre-trained using the coupled multi-physics model input-output and then updated with experimental data. Then, the model parameters of this pre-trained network are updated using the experimental data.

An alternative approach, denoted as $\mathbf{\rm GP^{\rm MF}}$, pursues the second approach of the second PIML strategy, i.e., it pre-trains a GP surrogate model with data generated from the physics model, then improves the surrogate using experimental data. Consider a physics model $G(\boldsymbol{\cdot})$ that maps input variables $\mathbf{X}$ and model parameters $\boldsymbol{\theta_m}$ to the numerical model output $\mathbf{Y}_{m}$:
\begin{equation}
\mathbf{Y}_{m}(\mathbf{X}) = G\big(\mathbf{X};\ \boldsymbol{\theta}_m(\mathbf{X})\big).
\end{equation}
Let $n_{D}$ be the number of collected observation data $\mathbf{Y_{\mathrm{obs}}}$ from experiments with input variable settings $\mathbf{x}^{(1)}, ..., \mathbf{x}^{(n_{D})}$, where $\mathbf{x}^{(i)}$ is the input variable setting for the $i$th experiment. The physics model prediction is inaccurate due to missing physics or due to other approximations. Thus, a model discrepancy term $\boldsymbol\delta(\mathbf{X})$ as a function of model inputs is introduced to capture the difference between $\mathbf{Y}_{\rm true}$ and $\mathbf{Y}_{m}$~\cite{kennedy2001bayesian}:
\begin{equation}
    \mathbf{Y}_{\rm true}(\mathbf{X}) = \mathbf{Y_{\mathrm{m}}}(\mathbf{X}) + \boldsymbol\delta(\mathbf{X}).
    \label{eq:modelerror}
\end{equation}

The experimental observations $\mathbf{Y}_{\rm obs}$ can be described in terms of the true system response $\mathbf{Y}_{\rm true}$ and the corresponding observation errors $\epsilon_{\rm obs}(\mathbf{X})$ as
\begin{equation}
    \mathbf{Y}_{\rm obs}(\mathbf{X}) = \mathbf{Y_{\mathrm{true}}}(\mathbf{X}) + \epsilon_{\rm obs}(\mathbf{X}).
    \label{eq:true_response}
\end{equation}
Combining Eqs.~\eqref{eq:modelerror} and~\eqref{eq:true_response},
\begin{equation}
    \mathbf{Y}_{\rm obs}(\mathbf{X}) - \epsilon_{\rm obs}(\mathbf{X}) = \mathbf{Y}_{m}(\mathbf{X}) + \boldsymbol\delta(\mathbf{X})=G\big(\mathbf{X};\ \boldsymbol{\theta}_m(\mathbf{X})\big) + \boldsymbol\delta(\mathbf{X}).
    \label{eq:combined}
\end{equation}

When the physics model is computationally expensive, it is replaced by a cheaper surrogate model. In Model 3, a GP surrogate model is used to approximate the original physics model. The accuracy of the surrogate model prediction depends on the quality and quantity of the training data generated by the original physics model. The surrogate model error ($\epsilon_{\delta}(\mathbf{X})$) can be incorporated as follows:
\begin{equation}
    \mathbf{Y}_{m}(\mathbf{X}) = \mathbf{\hat{Y}}_{m}(\mathbf{X}) + \epsilon_{\delta}(\mathbf{X}),
    \label{eq:sm_error}
\end{equation}
where $\mathbf{\hat{Y}}_{m}$ is the surrogate model prediction.

A common approach to estimate the discrepancy term $\boldsymbol\delta(\mathbf{X})$ is the one formulated by Kennedy and O’Hagan~\cite{kennedy2001bayesian}, which is applicable in the context of Bayesian calibration. In that case, physics model parameters are sought to be calibrated, and a discrepancy term is added in the calibration equation. The discrepancy term can be expressed in multiple ways, such as constant, Gaussian random variable with unknown parameters (either input-dependent or not), or Gaussian process (either stationary or non-stationary)~\cite{ling2014selection}. The hyperparameters of the discrepancy term are then estimated along with the physics model parameters using Bayesian calibration~\cite{berkcan_Optimization}.

However, the situation considered here is much simpler. There is no calibration of the physics model parameters here; only the discrepancy term is needed. (In other words, the physics model parameters are already established). In that case, the model discrepancy can be evaluated for different input values of experimental tests and realizations of observation errors as follows:
\begin{equation}
    \boldsymbol{\delta}(\mathbf{X})  = \mathbf{Y_{\mathrm{obs}}}(\mathbf{X}) - \epsilon_{\rm obs}(\mathbf{X}) - \mathbf{\hat{Y}}_{m}(\mathbf{X})-\epsilon_{\delta}(\mathbf{X}).
    \label{eq:modelFormError}
\end{equation}

Moving the surrogate model error $\epsilon_{\delta}(\mathbf{X})$ to the left-hand side, we can express the difference between the actual response and GP model prediction as
\begin{equation}
    \boldsymbol{\hat{\delta}}(\mathbf{X})  = \boldsymbol{\delta}(\mathbf{X})  + \epsilon_{\delta}(\mathbf{X}) = \mathbf{Y_{\mathrm{obs}}}(\mathbf{X}) - \epsilon_{\rm obs}(\mathbf{X}) - \mathbf{\hat{Y}}_{m}(\mathbf{X}).
    \label{eq:modelFormErrorupd}
\end{equation}

In this work, a second GP model is trained for $\boldsymbol{\hat{\delta}}(\mathbf{X})$ in terms of the inputs. Thus two GP models are trained in Model 3. The first GP model is constructed using the physics model input-output data, and predicts $\mathbf{\hat{Y}}_{m}$. The second GP model is constructed using the experimental data and the corresponding surrogate model predictions, and predicts $\boldsymbol{\hat{\delta}}$ (the difference between the surrogate model prediction and actual system response).

The GP model for the model discrepancy captures the combined contribution of measurement error, physics and surrogate model errors for a given prediction. Thus, the predictions of the first GP model (pre-trained) are corrected with the second GP model predictions ($\boldsymbol{\hat{\delta}}$) representing the model discrepancy term and can be written as
\begin{equation}
    \mathbf{\hat{Y}}(\mathbf{X}) =  \mathbf{\hat{Y}}_{m}(\mathbf{X}) + \boldsymbol{\hat{\delta}}.
    \label{eq:pred}
\end{equation}


Model 4, denoted as $\mathbf{\rm GP^{\rm upd, \mathcal{L}_{\rm phy}}}$, combines both PIML strategies for GP, where the optimized model parameters of the pre-trained model based on the physics model input-output are used as the initial values. These model parameters are updated using the experimental data by minimizing the augmented loss function shown in Eq.~\eqref{eq:lossfunc_gp} which consists of both the training loss function and the physics constraint loss terms.

\subsubsection{Implementation of PIML in DNN models}\label{Sec:DNN_PIML}
In Model 5, denoted simply as $\mathbf{\rm DNN}$, a deep neural network is trained using only experimental data. In order to train the model, an optimization algorithm is used to find a set of model parameters (weights and biases) that best map inputs to outputs. The number of epochs, which is the number of complete passes through a batch of training dataset, layers and neurons needs to be optimized to reduce cost and improve model accuracy. The model accuracy is evaluated using a validation dataset not used for training; the number of epochs is gradually increased until the accuracy improvement is insignificant. The accuracy and generalization performance of the DNN are also affected by the dropout rate used in training the model; the dropout concept will be discussed in Section~\ref{sec:BNN} below, and the selection of the dropout rate will be discussed in Section~\ref{Sec:Results}.

Model 6, denoted as $\mathbf{\rm DNN^{\mathcal{L}_{\rm phy}}}$, extends Model 5 by implementing the first PIML strategy, i.e., physical knowledge related to the physical process is enforced through constraints within the loss function of the DNN, as shown in Eq.~\eqref{eq:lossfunc}. The physics-based loss function terms are evaluated for given experimental inputs at every optimization iteration during the training of $\mathbf{\rm DNN^{\mathcal{L}_{\rm phy}}}$; this makes the optimization process slower during the training. In particular, the multipliers in the physics constraint penalty terms ($\lambda^{\rm DNN}_{\rm phy}$) affect the training speed; a higher value of the multiplier makes the penalty stronger (i.e., the physics constraints more stringent), thus requiring an increased number of iterations to converge to the optimum solution.

Model 7, denoted as $\mathbf{\rm DNN^{\rm upd}}$, pursues the second PIML strategy, where a DNN model is pre-trained using the coupled multi-physics model input-output and then updated with experimental data. The pre-trained model is first trained with physics model input-output data consisting of input combinations over a range of values. Then, the weights and biases (model parameters) of this pre-trained network are updated using the experimental data.


Model 8, denoted as $\mathbf{\rm DNN^{\rm upd, \mathcal{L}_{\rm phy}}}$, combines both PIML strategies for DNN, where the optimized model parameters of the pre-trained model based on the physics model input-output are used as the initial values. These model parameters are updated using the experimental data by minimizing the augmented loss function shown in Eq.~\eqref{eq:lossfnc_dnn} which consists of both the training loss function and the physics constraint loss terms. Similar to Model 6, the inclusion of physics constraints makes it slower for the optimization to converge to optimal model parameter values.

\subsection{Variance of GP and DNN prediction}\label{sec:BNN}
In general, every surrogate model has uncertainty in prediction, whether acknowledged or not. In the GP models, the prediction at a given input is expressed by a normal distribution with a mean and variance. In order to quantify the uncertainty in the GP prediction, we can sample multiple realizations of the Gaussian process. Note that this only captures the variance of the GP prediction, not the bias, which can be evaluated by comparing against validation data.

In the DNN models, the estimates of the model parameters (neuron weights $\mathbf{w}$) have uncertainty, and this uncertainty depends on the available training data. When the neural network parameters are represented using distributions (to reflect the epistemic uncertainty) instead of deterministic values, the model is referred to as a Bayesian neural network (BNN)~\cite{denker1991transforming,mackay1992practical,neal2012bayesian}. In this Bayesian context, the model parameter uncertainty is first described using a prior distribution $p(\mathbf{w})$, and the  likelihood function is $p(\mathbf{Y} | \mathbf{X}, \mathbf{w})$. Following Bayes' theorem, a posterior distribution over the model parameters given the training data $\{\mathbf{X}_T,\mathbf{Y}_T\} = \{\{\mathbf{x}_1, ...,\mathbf{x}_N\}, \{\{\mathbf{y}_1, ...,\mathbf{y}_N\}\}$ is defined by
\begin{equation}
    p(\mathbf{w} | \mathbf{X}, \mathbf{Y}) = \frac{p(\mathbf{Y} | \mathbf{X}, \mathbf{w}) p( \mathbf{w})}{p(\mathbf{Y} | \mathbf{X})}.
\end{equation}

In this context, the predictive distribution of the model outputs for a given input $\hat{\mathbf{X}}$ is given by:
\begin{equation}\label{eq:pred_post_dnn}
    p(\mathbf{\hat{\mathbf{Y}}} | \hat{\mathbf{X}}, \mathbf{X}_T, \mathbf{Y}_T) = \int_{\Omega} p(\hat{\mathbf{Y}} | \hat{\mathbf{X}}, \mathbf{w}) p(\mathbf{w} | \mathbf{X}_T, \mathbf{Y}_T) d\mathbf{w}.
\end{equation}
The posterior distribution of model parameters $p(\mathbf{w} | \mathbf{X}_T, \mathbf{Y}_T)$ is challenging to evaluate over the entire parameter space $\Omega$ due to the high dimensionality of $\Omega$ in a DNN model, and the highly non-linear behavior caused by the non-linear activation functions and their combinations across multiple hidden layers. Therefore, different approximate inference techniques can be considered to infer the posterior distribution $p(\mathbf{w} | \mathbf{X}_T, \mathbf{Y}_T)$~\cite{blundell2015weight,graves2011practical,hernandez2016black,gal2015bayesian}. One such approximation is variational inference, which fits a simple and tractable distribution $q_{\theta}(\mathbf{w})$ to the posterior, parametrized by a variational parameter $\theta$~\cite{blundell2015weight}. This approximates the intractable problem by optimizing the parameters of $q_{\theta}(\mathbf{w})$. The quality of the variational inference can be assessed by the Kullback-Leibler (KL) divergence between the approximate distribution $q_{\theta}(\mathbf{w})$ and the true model posterior $p(\mathbf{w} |  \mathbf{X}_T, \mathbf{Y}_T)$. 

The term \emph{dropout} refers to randomly dropping out neurons (along with their connections) with a given dropout rate during the training phase in a neural network. Dropout is a common regularization approach in neural network training, which prevents over-fitting and reduces generalization error. A Monte Carlo (MC) dropout technique has been developed in recent years in the context of Bayesian neural networks~\cite{gal2016dropout}, which has been shown to be equivalent to performing approximate variational inference. In MC dropout, dropout is not only applied while training a model but also during prediction. Randomly chosen neurons are temporarily removed from the network along with their connections. Next, the gradients of neuron weights are calculated on a sub-neural network for each training data and these gradients are then averaged over the training sets to obtain the weights for the overall network. A Bayesian neural network with MC dropout generates random samples following a binomial distribution (0 or 1) for each neuron in the input and hidden layers during prediction. The neuron that takes the value $0$ is dropped with probability $p_d$. The outputs of the network are predicted using the collection of generated random samples from the posterior predictive distribution and the uncertainty in the prediction of a new data is quantified with the trained network. Thus the MC dropout strategy provides an efficient way of Bayesian inference to quantify the model prediction variance, and can be applied to a variety of neural networks, such as feedforward neural networks, convolutional neural networks, and recurrent neural networks ~\cite{zhang2020bayesian}.

The sensitivity estimate results depend on the dropout rate. The main reason for this is that the model is regularized and it underfits the data as it is over-regularized. Further, both the accuracy and uncertainty of the sensitivity estimates also depend on the number of training epochs, and the architecture of the network. If the model is not fully trained, it will also result in underfitting, leading to larger bias and variance in the prediction. 

\subsection{Sobol' indices computation with model uncertainty}
This section discusses the incorporation of ML model prediction variance within the estimation of Sobol' indices using the GP and DNN models.

When the training data is noise-free, the GP predictions at the training points have zero variance and at other points the variance is non-zero. The prediction at any point is given by a normal distribution with a mean and variance. This prediction uncertainty can be captured by sampling multiple realizations of the GP model, which can then be used in GSA. The model uncertainty pertaining to the GP model is propagated to the Sobol’ index calculations using the following estimator~(see \cite{le2014bayesian}):
\begin{align}\label{eq:sens_est_gp}
    S^{\rm GP}_{i} = \frac{V_{X^i}\big(E_{\mathbf{X}}\big[\mathbf{y}_P(\mathbf{X})|X^{i}\big]\big)}{V\big(\mathbf{y}_P(\mathbf{X})\big)} \approx \frac{\frac{1}{m}\sum_{k=1}^m \mathbf{y}_P(\mathbf{X}_k) \mathbf{y}_P(X_k^{i}) - \frac{1}{m}\sum_{k=1}^m \mathbf{y}_P(\mathbf{X}_k) \sum_{k=1}^m \mathbf{y}_P(X_k^{i})}{\frac{1}{m} \sum_{k=1}^m \mathbf{y}_P(\mathbf{X}_k)^2 - \big[\frac{1}{m}\sum_{k=1}^m \mathbf{y}_P(\mathbf{X}_k)\big]^2}
\end{align}
where $\mathbf{y}_P(\mathbf{X})$ is a realization of the predictive distribution shown in Eq.~\eqref{eq:predictiveDist} trained using experimental data, and $\mathbf{X}_k$ and $X_k^{i}$ are the $k$th samples of the random vectors  $\mathbf{X}$ and $\mathbf{X}^i$.



The distribution of $S^{\rm GP}_{i}$ can be computed by sampling $N_Z$ realizations from the Gaussian predictive distribution $\mathbf{Y}_P(\mathbf{X})$ numerically using Algorithm~\ref{alg:1}.
\begin{algorithm}[H]
	\caption{Estimation of the distribution of $S^{\rm GP}_{i}$ using GP models.}
	\label{alg:1}
	\begin{algorithmic}[1]
	\State Generate two samples $\mathbf{X}_k$ and $X_k^{i}$ ${(k=1,\ldots,m)}$ of the random vectors $\mathbf{X}$ and $X^{i}$.
		\For {$p=1,2,\ldots,N_Z$}
			\State Sample a realization $\mathbf{y}_P(\mathbf{x})$ of $\mathbf{Y}_P(\mathbf{X})$ with $\mathbf{x}=\{(x_k)_{k=1,\ldots,m}, (x_k^{i})_{k=1,\ldots,m}\}$.
			\State Compute $\hat{S}^{\rm GP}_{i,p}$ using Eq.~\eqref{eq:sens_est_gp}.
		\EndFor
		
	\Return $(\hat{S}^{\rm GP}_{i,p})_{p=1,2,\ldots,N_Z}$.
	\end{algorithmic} 
\end{algorithm}

The output of Algorithm~\ref{alg:1} $(\hat{S}^{\rm GP}_{i,p})_{p=1,2,\ldots,N_Z}$ is a sample of size $N_Z$, where $m$ is the number of Monte Carlo samples. Thus, the mean and variance of sensitivity estimates obtained using the GP models are defined as follows, respectively:
\begin{align}\label{eq:mu_var_gp}
    & \mu_{S^{\rm GP}_{i}} = \frac{1}{N_Z} \sum_{p=1}^{N_Z} \hat{S}^{\rm GP}_{i,p}, \\ \notag
    & \sigma^2_{S^{\rm GP}_{i}} = \frac{1}{N_Z} \sum_{p=1}^{N_Z} (\hat{S}^{\rm GP}_{i,p} - \mu_{S^{\rm GP}_{i}})^2.
\end{align}

A similar approach can be implemented in the DNN models with the use of MC dropout (with a chosen dropout rate). However, in contrast to the GP models, where we sample from a multivariate normal distribution to quantify the uncertainty in the Sobol’ index estimates, the sampling implementation is different in the DNN models. In the DNN models, we randomly set units of the network to zero and generate predictions using the remaining units of the network as shown in Algorithm~\ref{alg:2}. The neuron weights can be drawn from the approximate posterior $\hat{\mathbf{w}}\sim q_{\theta}(\mathbf{w})$ to obtain the model outputs of a DNN with MC dropout denoted as $\mathcal{G}^{\hat{w}}(\mathbf{X}) = \hat{\mathbf{Y}}$, where $\hat{\mathbf{Y}}$ is the predictive mean,

The predictive posterior given in Eq.~\ref{eq:pred_post_dnn} can be defined as follows:
\begin{equation}
    p(\mathbf{\hat{\mathbf{Y}}} | \hat{\mathbf{X}}, \mathbf{X}_T, \mathbf{Y}_T) = \mathcal{N}(\hat{\mathbf{Y}}; \mathcal{G}^{\hat{w}}(\mathbf{X}), \sigma^2 \mathbf{I})
\end{equation}
where $\sigma^2=(2N\lambda)/((1-p_d)l^2)$ is the noise term~\cite{gal2016dropout}, $l$ being the prior length-scale, and $\lambda$ is the regularization strength used in typical loss functions. Dropout can be interpreted as a variational Bayesian approximation and the minimization objective is defined by~\cite{jordan1999introduction}
\begin{equation}
    \mathcal{L}(\theta, p_d)=-\frac{1}{N}\sum_{i=1}^{N}\mathrm{log}\ p(\hat{\mathbf{Y_i}}|\mathcal{G}^{\hat{w}}(\mathbf{X_i})) + \frac{1-p_d}{2N}||\theta||^2
\end{equation}
where $\theta$ is the set of distribution's parameters to be optimized (i.e., weights of the network).

The predictive mean and predictive uncertainty are estimated by collecting the results of stochastic forward passes through the model. The mean prediction of the model with $N_d$ samples can be approximated by
\begin{equation}\label{eq:BNNpredmean}
    \mathbb{E}(\mathbf{Y}) \approx \frac{1}{N_d} \sum_{t=1}^{N_d} \mathcal{G}^{\hat{w}}(\mathbf{X}),
\end{equation}
and the variance of the prediction is estimated by
\begin{equation}\label{eq:BNNpredvar}
    \text{Var}(\mathbf{Y}) \approx \sigma^2 + \frac{1}{N_d} \sum_{t=1}^{N_d} \mathcal{G}^{\hat{w}}(\mathbf{X})^T\mathcal{G}^{\hat{w}}(\mathbf{X}) - \mathbb{E}(\mathbf{Y})^T\mathbb{E}(\mathbf{Y}).
\end{equation}

Similar to Eq. ~\eqref{eq:sens_est_gp}, the uncertainty in the DNN model can be propagated to the sensitivity calculations using the following estimator:
\begin{align}\label{eq:sens_est_bnn}
    S^{\rm DNN}_{i} = \frac{\frac{1}{m}\sum_{k=1}^m \mathcal{G}^{\hat{w}}(\mathbf{X}_k)\mathcal{G}^{\hat{w}}(\mathbf{X}_k^{\prime}) - \frac{1}{m}\sum_{k=1}^m \mathcal{G}^{\hat{w}}(\mathbf{X}_k) \sum_{k=1}^m \mathcal{G}^{\hat{w}}(\mathbf{X}_k^{\prime})}{\frac{1}{m} \sum_{k=1}^m \mathcal{G}^{\hat{w}}(\mathbf{X}_k)^2 - \frac{1}{m}\sum_{k=1}^m (\mathcal{G}^{\hat{w}}(\mathbf{X}_k))^2},
\end{align}
where $\mathcal{G}^{\hat{w}}(\mathbf{X})$ denotes the deep neural network (DNN) with MC dropout and all the other terms have the same definition as Eq.~\eqref{eq:sens_est_gp}. We note that the model uncertainty is propagated to the calculation of the sensitivity estimates by directly using the results based on stochastic forward passes through a dropout-reduced DNN instead of the predictive mean and predictive uncertainty. 
\begin{algorithm}[H]
	\caption{Estimation of the distribution of $S^{\rm DNN}_{m}$ using DNN models with MC dropout.}
	\label{alg:2}
	\begin{algorithmic}[1]
	\State Generate samples $\mathbf{X}_k$ and $\mathbf{X}_k^{\prime}$ ${(k=1,\ldots,m)}$ of the random vectors $\mathbf{X}$ and $\mathbf{X}^{\prime}$.
		\For {$p=1,2,\ldots,N_{d}$}
			\State Perform a stochastic forward pass through the network $\mathcal{G}^{\hat{w}}(\mathbf{X})$ using MC dropout and calculate the model prediction $\mathbf{\hat{Y}}=\mathcal{G}^{\hat{w}}(\mathbf{X})$.
			\State Compute $\hat{S}^{\rm DNN}_{m,p}$ using Eq.~\eqref{eq:sens_est_bnn}.
		\EndFor
		
	\Return $(\hat{S}^{\rm DNN}_{m,p})_{p=1,2,\ldots,N_{d}}$.
	\end{algorithmic} 
\end{algorithm}

The output of Algorithm~\ref{alg:2} $(\hat{S}^{\rm DNN}_{m,p})_{p=1,2,\ldots,N_d}$ is a sample of size $N_d$. Thus, the mean and variance of sensitivity estimates obtained using DNN models are defined as follows, respectively:
\begin{align}\label{eq:mu_var_bnn}
    & \mu_{S^{\rm DNN}_{m}} = \frac{1}{N_{d}} \sum_{p=1}^{N_{d}} \hat{S}^{\rm DNN}_{m,p}, \\ \notag
    & \sigma^2_{S^{\rm DNN}_{m}} = \frac{1}{N_{d}} \sum_{p=1}^{N_{d}} (\hat{S}^{\rm DNN}_{m,p} - \mu_{S^{\rm DNN}_{m}})^2.
\end{align}

In summary, two types of PIML strategies are proposed in this section and implemented in two types of ML models (GP and DNN), in order to evaluate the accuracy and uncertainty of the sensitivity estimates in GSA. Eight different PIML models are developed by leveraging the two PIML strategies. The accuracy of these models can be assessed by comparison against validation data, whereas the variance of the sensitivity estimates can be quantified using Algorithms~\ref{alg:1} and~\ref{alg:2} and Eqs.~\ref{eq:sens_est_gp} to~\ref{eq:mu_var_bnn}. The different models have different training strategies and different numbers of parameters, both of which will affect the accuracy and uncertainty of the sensitivity estimates. These differences are assessed in detail in the next section.

\section{Numerical illustration}\label{Sec:Results}
\subsection{Illustrative example 1}
\subsubsection{Problem setup}
An additive manufacturing application is used to illustrate the proposed PIML models for GSA and compare their performance. A fused filament fabrication (FFF) process is considered; commercial material Ultimaker Black Acrylonitrile butadiene styrene (ABS) is extruded from an Ultimaker 2 Extended+ 3D printer to manufacture parts with unidirectionally aligned filaments, and the porosity of the manufactured part is measured. FFF is a widely used additive manufacturing (AM) process due to its easy operation, low cost, and suitability for complex geometries. As the molten filament is deposited layer upon layer through a nozzle, it cools down, solidifies and bonds with the adjacent filaments. Rectangular specimens of length 35 mm, width 12 mm, and thickness 4.2 mm are manufactured for the ABS amorphous polymer, with constant filament height, width and length (0.7, 0.8, and 35 mm, respectively).

The output QoI is the porosity of the printed part, and the inputs are two process parameters, namely nozzle temperature and speed. The porosity of an FFF part is dependent on the temperature history at the interfaces between filaments. Thus, it is important to predict the temperature evolution of filaments for estimating the final mesostructure of the printed part. The analytical solution proposed by Costa et al.~\cite{costa2017estimation} for transient heat transfer during the printing process in FFF is used to predict the temperature evolution of filaments. A physics-based sintering model is developed, which considers realistic filament geometry, and allows the filament geometry to change during the printing process~\cite{berkcan2020piml}. This model is used to predict the porosity of the FFF part using the temperature evolution of filaments, material properties, part geometry, and process parameters as inputs. Thus the mapping from input to output is a multi-physics model, i.e., models of two physical phenomena (heat transfer and sintering) are combined to predict the porosity given the values of two process parameters: nozzle temperature and nozzle speed.

\begin{figure}[!ht]
\centering
\includegraphics[width=0.6\textwidth]{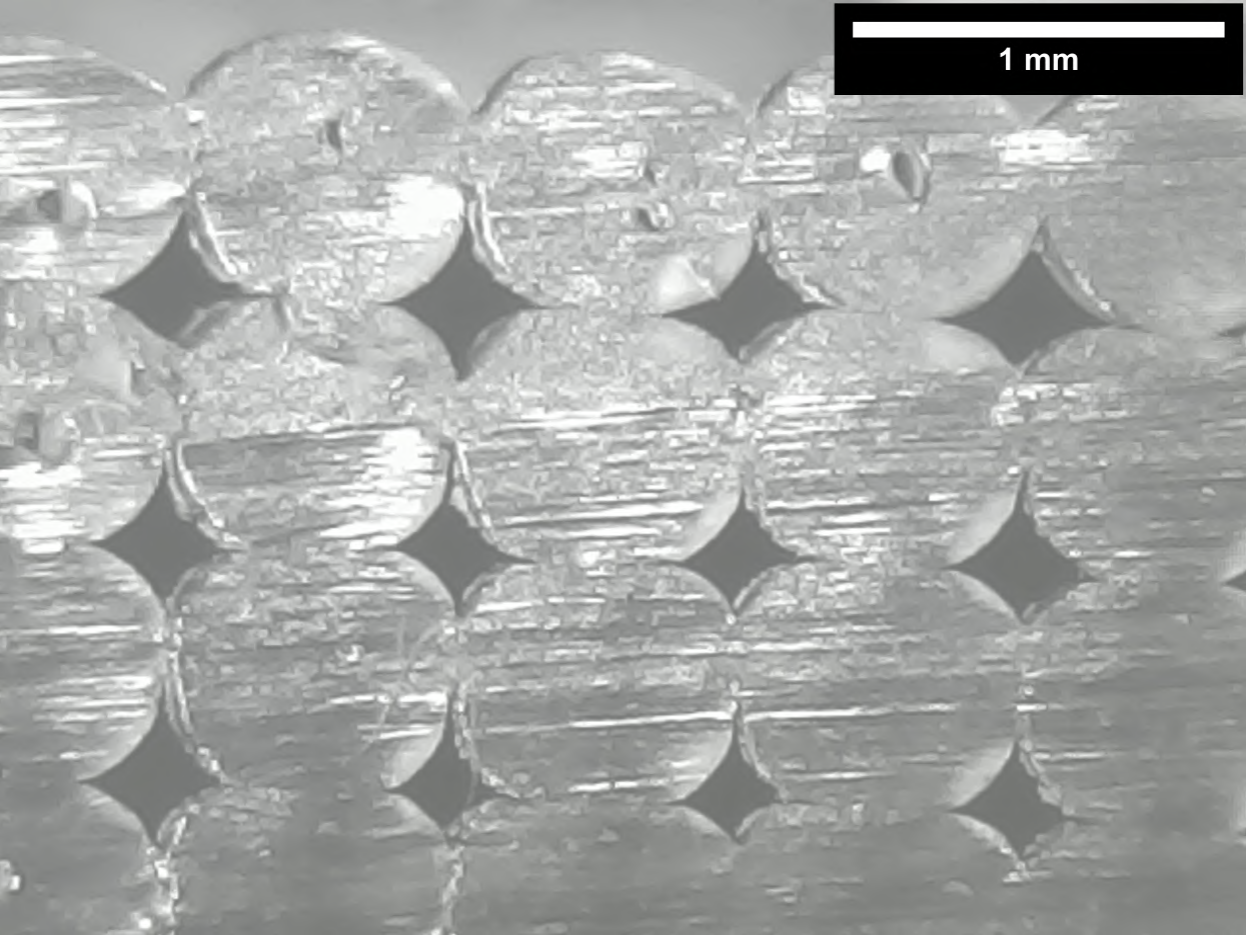}
\caption{Cross-sectional geometry of an FFF specimen printed with nozzle temperature 240$^{\circ}$C and speed 42\ mm/s.}
\label{fig:CrossSection}
\end{figure}

The statistical properties of the QoI were observed to have negligible variability along the length of the specimens; therefore only the porosity measurements taken at the midpoint cross-section (see Fig.~\ref{fig:CrossSection}) are discussed here. These measurements were based on microscopy images processed through the ImageJ software~\cite{schneider2012nih}.  Filaments were extruded through a nozzle with 0.8 mm diameter. The build plate temperature was constant and set to $110^{\circ}$C. Using Latin hypercube sampling, 39 sets of process parameters $\mathbf{X}$ were generated, and experiments were conducted at these 39 values. The ranges considered for the two process variables were: nozzle temperature $T$: (210$^{\circ}C$ - $260 ^{\circ}C$), and nozzle speed $S$: (15 mm/s - 46 mm/s).

\subsubsection{Training details of the ML models}

The basic ML models, namely Model 1 (for GP) and Model 5 (for DNN) are simply trained with the 39 sets of process inputs (temperature and speed) and output (porosity).

In the training of Model 2 ($\mathbf{\rm GP^{\mathcal{L}_{\rm phy}}}$) and Model 6 ($\mathbf{\rm DNN^{\mathcal{L}_{\rm phy}}}$),  we impose two physics constraints (i.e., two separate loss function terms, $\mathcal{L}_{{\rm phy}, k}(\mathbf{\hat{Y}})$, where $k=\{1,2\}$ and $\mathbf{\hat{Y}}$ is the porosity prediction). The corresponding loss function terms are defined as
\begin{align}\label{eq:phyloss}
    & {}\mathcal{L}_{{\rm phy}, 1}(\mathbf{\hat{Y}}) = \frac{1}{N} \sum_{i=1}^N \rm{ReLU}(-\mathnormal{\hat{Y}}_{i}), \notag\\
    & {}\mathcal{L}_{{\rm phy}, 2}(\mathbf{\hat{Y}}) = \frac{1}{N} \sum_{i=1}^N \rm{ReLU}(\mathnormal{\hat{Y}}_{i}-\phi_{0,i}),
\end{align}
considering physics violations related to the porosity in all the $N$ samples. In the first loss function, a negative value of porosity is treated as a physics violation. The second loss function penalizes the model when the predicted final porosity  $\mathnormal{\hat{Y}}_{i}$ is greater than the initial porosity $\phi_{0,i}$ of the $i$th part. This is based on the physics knowledge that the total void area decreases as the bond formation takes place. Thus, the porosity predictions are ensured to be physically meaningful with the inclusion of these physics-based penalty terms.

The overall “loss” function of the GP model is
\begin{align}\label{eq:lossfnc_gp}
    \mathcal{L_{\mathrm{\overline{GP}}}} = \mathcal{L}_{\rm GP} -  \lambda^{\rm GP}_{\mathrm{phy,1}}\mathcal{L}_{\rm phy,1}(\mathbf{\hat{Y}}) - \lambda^{\rm GP}_{\mathrm{phy,2}}\mathcal{L}_{\rm phy,2}(\mathbf{\hat{Y}}).
\end{align}
Note that the GP model parameters are obtained by maximizing the above function. 

The overall loss function of the DNN model is
\begin{align}\label{eq:lossfuncdnn}
    \mathcal{L_{\mathrm{\overline{DNN}}}} = \mathcal{L}_{\rm DNN} + \lambda^{\rm DNN}_{\mathrm{phy,1}}\mathcal{L}_{\rm phy,1}(\mathbf{\hat{Y}}) + \lambda^{\rm DNN}_{\mathrm{phy,2}}\mathcal{L}_{\rm phy,2}(\mathbf{\hat{Y}}).
\end{align}
Note that the DNN model parameters are obtained by minimizing the above function.

In Model 3 ($\mathbf{\rm GP^{\rm upd}}$) and Model 7 ($\mathbf{\rm DNN^{\rm upd}}$), the ML models are pre-trained using the multi-physics model input-output. The pre-trained ML models are then updated using the experimental data. The training data for pre-training consists of 1310 input parameter combinations over a range of experimental values, i.e., (210$^{\circ}C$  $\le T \le$ $260 ^{\circ}C$,  15 mm/s $\le S\le$ 46 mm/s ). Note that there are only 39 physical experiments available; this is one of the advantages of the pre-training/updating strategy, where the pre-training can be over a much larger set of input combinations, thus improving the generalization performance of the updated model. The input data are normalized prior to the training of the ML models (i.e., the output quantity porosity is dimensionless and between 0 and 1).

Model 4 ($\mathbf{\rm GP^{\rm upd, \mathcal{L}_{\rm phy}}}$) combines both PIML strategies for GP, and consists of two GP models: (i) the first GP model is trained using the physics model input-output samples consisting of 1310 input parameter combinations; and (ii) the second GP model is built for the discrepancy between the first GP model prediction and the actual system response using the experimental data, by maximizing the function shown in Eq.~\eqref{eq:lossfnc_gp} to optimize the hyperparameters of the second GP model.

Model 8 $\mathbf{\rm DNN^{\rm upd, \mathcal{L}_{\rm phy}}}$, which is a combination of the two PIML strategies, uses the DNN model parameters trained using the physics model input-output as the initial values. Then, during the updating phase with the experimental data, these parameters are updated by minimizing the loss function shown in Eq.~\eqref{eq:lossfuncdnn}.

The four GP models (Models 1 to 4) were implemented using Python. The optimization of the hyperparameters were performed using the scikit-optimize package. The multipliers of the physics constraint terms of models 2 ($\mathbf{\rm GP^{\mathcal{L}_{\rm phy}}}$) and 4 ($\mathbf{\rm GP^{\rm upd, \mathcal{L}_{\rm phy}}}$) were chosen as $(\lambda^{\rm GP}_{\rm phy,1},\lambda^{\rm GP}_{\rm phy,2})= (50, 50)$ based on a cross-validation test. The Automatic Relevance Determination (ARD) squared exponential function~\cite{williams2006gaussian} was used as the covariance function for all the GP models.

The four DNN models (Models 5 to 8) were implemented using the Keras package~\cite{chollet2015keras} with Tensorflow in the backend. The hyperparameters of each model were tuned with grid search and the multipliers of the physics constraint terms in Model 6 ($\mathbf{\rm DNN^{\mathcal{L}_{\rm phy}}}$) and Model 8 ($\mathbf{\rm DNN^{\rm upd, \mathcal{L}_{\rm phy}}}$) were chosen as $(\lambda^{\rm DNN}_{\rm phy,1},\lambda^{\rm DNN}_{\rm phy,2})= (0.01, 0.01)$ based on a cross-validation test. Fully-connected DNN models with 2 hidden layers and 5 neurons in each hidden layer were constructed. The Rectified Linear Unit (ReLU) activation function and Adam optimizer were used to perform stochastic gradient descent for 300 epochs in learning the model parameters. The dropout rate for the DNN models was chosen to be 0.05, for reasons as explained below.

\subsubsection{Comparison of computational effort}
The computational costs of different models for training and estimation of Sobol’ indices based on 5000 MC samples with a fixed number of experimental data ($n=39$) is given in Table.~\ref{table:Cost}. Among the GP models, the time it takes for training as well as computation of Sobol’ indices using Models 2-4 is significantly greater than Model 1. The reason for the difference between the training time of $\mathbf{\rm GP}$ and $\mathbf{\rm GP^{\rm upd}}$ is the pre-training phase, where a large amount of physics input-output samples used. Whereas, the difference between the training time of $\mathbf{\rm GP}$ and $\mathbf{\rm GP^{\mathcal{L}_{\rm phy}}}$ is due to the inclusion of physics constraints, which makes it harder for the optimization to find optimal hyperparameters. Interestingly, the training time for the DNN models ranges from 20 to 55 sec on the same desktop computer as used for the GP model (Intel\textsuperscript{\tiny\textregistered} Xeon\textsuperscript{\tiny\textregistered} CPU E5-1660 v4$@$3.20GHz with 32 GB RAM and GPU NVIDIA Quadro K620 with 2 GB). Among the DNN models, the reasons for the increased training time of models 6-8 compared to Model 5 are the same as for the GP models. The Sobol’ index estimations based on 5000 samples take approximately 1-2 minutes for the DNN models (models 5-8), whereas the GP predictions take much longer because the covariance matrix needs to be stored and inverted.
\begin{table*}[!htb]
    \caption{Computational effort of eight models for training and estimation of first-order and total effect Sobol' indices using 5000 MC samples with $n=39$ number of observations.}
    \label{table:Cost}
\centering{%
    \begin{tabular*}{\linewidth}{@{\extracolsep{\fill}}l*{3}c@{\extracolsep{\fill}}}
    \toprule
            Models & Training [in minutes] & Sobol' indices calculation [in minutes] \\
    \midrule
            1. $\mathbf{\rm GP}$ & 1 & 3 \\
            2. $\mathbf{\rm GP^{\mathcal{L}_{\rm phy}}}$ & 2 & 5 \\
            3. $\mathbf{\rm GP^{\rm upd}}$ & 3 & 7 \\
            4. $\mathbf{\rm GP^{\rm upd, \mathcal{L}_{\rm phy}}}$ & 3 & 8 \\
            5. $\mathbf{\rm DNN}$ & .3 & 1  \\
            6. $\mathbf{\rm DNN^{\mathcal{L}_{\rm phy}}}$ & .7 & 2  \\
            7. $\mathbf{\rm DNN^{\rm upd}}$ & .5 & 2 \\
            8. $\mathbf{\rm DNN^{\rm upd, \mathcal{L}_{\rm phy}}}$ & 1 & 2  \\
    \bottomrule
    \end{tabular*}}%
\end{table*}

\subsubsection{Comparison of accuracy}\label{sec:accuracy}
In order to compare the accuracy of the eight different models, the models are trained with different amounts of experimental observations ($n=(5,10,15,20,30)$), and the remaining 9 observations are used to compute the errors; the root mean square error (RMSE) based on the 9 validation samples is reported as the accuracy measure for comparison. The mean and one standard deviation RMSE values for the four GP models and the four DNN models are shown in Tables~\ref{table:EffectOfTrainingData_GP} and~\ref{table:EffectOfTrainingData_DNN} respectively, for different values of $n$. To further validate the accuracy of the models, the data set is divided into two subsets for cross-validation; $k$-fold cross-validation is performed by splitting the data set into sets for model training and cross-validation, and these sets are selected randomly $k=10$ different times. The average cross-validation accuracy of the models over the 10 folds (random shuffles) is assessed by evaluating the average RMSE (see Table~\ref{table:AvgRMSE}).

The results in Tables~\ref{table:EffectOfTrainingData_GP},~\ref{table:EffectOfTrainingData_DNN} and~\ref{table:AvgRMSE} show that the use of PIML strategies in DNN models improves the performance, and the improvement is relatively larger as the amount of observed data $n$ gets smaller. For $n$=20 and $n$=30, the basic DNN model is as accurate as the DNN models with the PIML strategies; whereas for smaller values of $n$, the DNN models with the PIML strategies are significantly more accurate. This clearly indicates the benefit of PIML over basic ML for the DNN models. However, the GP models incorporating PIML strategies do not show significant improvement. This is because the GP models have a much smaller number of parameters compared to the DNN models, and the achievable accuracy of any ML model is constrained by the number of model parameters (in addition to model form).
\begin{table}
    \caption{Effect of different amounts of training data on the RMSE of the GP models.}
    \label{table:EffectOfTrainingData_GP}
    \resizebox{1\columnwidth}{!}{%
        \begin{tabular}{@{\extracolsep{\fill}}l*{5}c@{\extracolsep{\fill}}}
        \hline\noalign{\smallskip}
            Model& $n=5$ & $n=10$ & $n=15$ & $n=20$ & $n=30$\\
        \noalign{\smallskip}\hline\noalign{\smallskip}
            1. $\mathbf{\rm GP}$ & 0.020($\pm$0.004) & 0.026($\pm$0.005) & 0.021($\pm$0.001) & 0.021($\pm$0.002) & 0.020($\pm$0.001) \\
            2. $\mathbf{\rm GP^{\mathcal{L}_{\rm phy}}}$ & 0.019($\pm$0.004) & 0.026($\pm$0.004) & 0.024($\pm$0.004) & 0.021($\pm$0.002) & 0.019($\pm$0.001)\\
            3. $\mathbf{\rm GP^{\rm upd}}$ & 0.021($\pm$0.003) & 0.028($\pm$0.003) & 0.020($\pm$0.001) & 0.020($\pm$0.001) & 0.019($\pm$0.001)\\
            4. $\mathbf{\rm GP^{\rm upd, \mathcal{L}_{\rm phy}}}$ & 0.021($\pm$0.005) & 0.023($\pm$0.003) & 0.022($\pm$0.003) & 0.020($\pm$0.002) & 0.019($\pm$0.001)\\
        \noalign{\smallskip}\hline
        \end{tabular}
    }%
\end{table}

\begin{table}
    \caption{Effect of different amounts of training data on the RMSE of the DNN models.}
    \label{table:EffectOfTrainingData_DNN}
    \resizebox{1\columnwidth}{!}{%
        \begin{tabular}{@{\extracolsep{\fill}}l*{5}c@{\extracolsep{\fill}}}
        \hline\noalign{\smallskip}
            Model& $n=5$ & $n=10$ & $n=15$ & $n=20$ & $n=30$\\
        \noalign{\smallskip}\hline\noalign{\smallskip}
            5. $\mathbf{\rm DNN}$ & 0.027($\pm$0.007) & 0.017($\pm$0.009) & 0.019($\pm$0.006) & 0.013($\pm$0.003) & 0.013($\pm$0.003) \\
            6. $\mathbf{\rm DNN^{\mathcal{L}_{\rm phy}}}$ & 0.017($\pm$0.007) & 0.015($\pm$0.008) & 0.014($\pm$0.005) & 0.014($\pm$0.003) & 0.013($\pm$0.002)\\
            7. $\mathbf{\rm DNN^{\rm upd}}$ & 0.014($\pm$0.002) & 0.009($\pm$0.002) & 0.014($\pm$0.001) & 0.013($\pm$0.001) & 0.013($\pm$0.001)\\
            8. $\mathbf{\rm DNN^{\rm upd, \mathcal{L}_{\rm phy}}}$ & 0.014($\pm$0.002) & 0.009($\pm$0.001) & 0.014($\pm$0.001) & 0.014($\pm$0.001) & 0.013($\pm$0.001)\\
        \noalign{\smallskip}\hline
        \end{tabular}
    }%
\end{table}

\begin{table}
    \caption{Tenfold cross-validation average RMSE results of GP and DNN models}
    \label{table:AvgRMSE}
    \resizebox{1\columnwidth}{!}{%
        \begin{tabular}{@{\extracolsep{\fill}}l*{9}c@{\extracolsep{\fill}}}
        \hline\noalign{\smallskip}
            \multicolumn{9}{c}{Models} \\
            & 1. $\mathbf{\rm GP}$ & 2. $\mathbf{\rm GP^{\mathcal{L}_{\rm phy}}}$ & 3. $\mathbf{\rm GP^{\rm upd}}$ & 4. $\mathbf{\rm GP^{\rm upd, \mathcal{L}_{\rm phy}}}$ & 5. $\mathbf{\rm DNN}$ & 6. $\mathbf{\rm DNN^{\mathcal{L}_{\rm phy}}}$ & 7. $\mathbf{\rm DNN^{\rm upd}}$ & 8. $\mathbf{\rm DNN^{\rm upd, \mathcal{L}_{\rm phy}}}$ & \\
        \noalign{\smallskip}\hline\noalign{\smallskip}
            $n=39$ & 0.021($\pm$0.002) & 0.020($\pm$0.002) & 0.020($\pm$0.001) &  0.019($\pm$0.002) & 0.015($\pm$0.004) & 0.013($\pm$0.004) & 0.013($\pm$0.003) & 0.013($\pm$0.003)\\
        \noalign{\smallskip}\hline
        \end{tabular}
    }%
\end{table}

\begin{figure}[!ht]
	\centering
	\includegraphics[width=.5\textwidth]{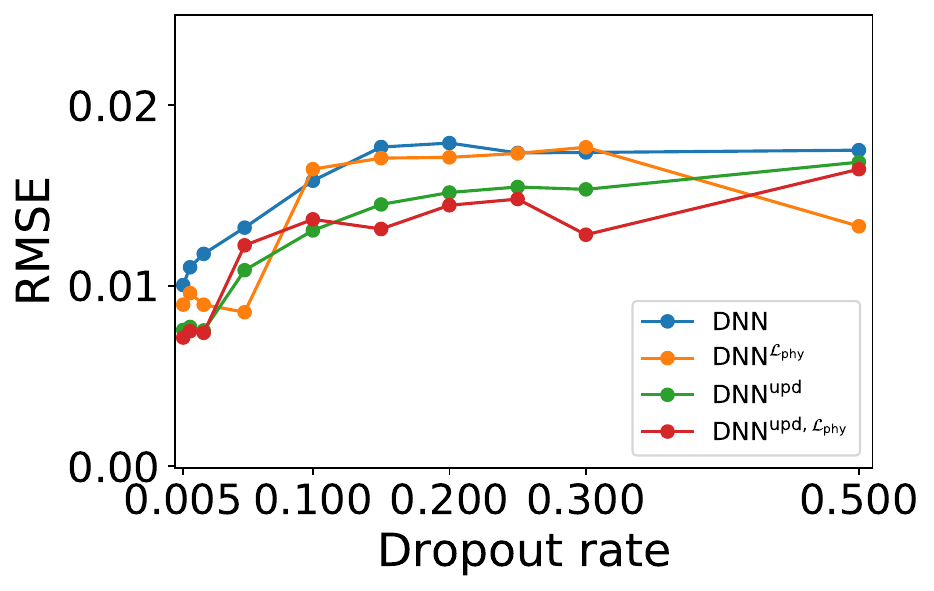}
	\caption{RMSE values of DNN models with varying dropout rates.}
	\label{fig:Dropout_vs_RMSE}
\end{figure}

In addition to the increased number of parameters, the DNN models have additional advantages in terms of training epochs and dropout rate that affect the prediction accuracy. As mentioned earlier, the optimum number of training epochs was found to be 300. Regarding dropout rate, the RMSE values (based on 9 validation samples) of DNN models with MC dropout for different dropout rates are shown in Fig.~\ref{fig:Dropout_vs_RMSE}. The lowest RMSE values for all four models are obtained between 0.005-0.05. On the other hand, the Sobol’ index estimates were found to be similar within this range of dropout rate, for each of the four DNN models. Therefore, a dropout rate of 0.05 was chosen for the reporting of GSA results, since a higher dropout rate results in smaller sub-networks, therefore a smaller number of parameters and faster training. By comparing Tables~\ref{table:EffectOfTrainingData_GP} and~\ref{table:EffectOfTrainingData_DNN}, it is seen that at the dropout rate of 0.05, the DNN models are more accurate compared to the GP models.

\begin{figure}[!htb]
\centering
  \begin{tabular}{@{}cc@{}}
    \includegraphics[width=.49\textwidth]{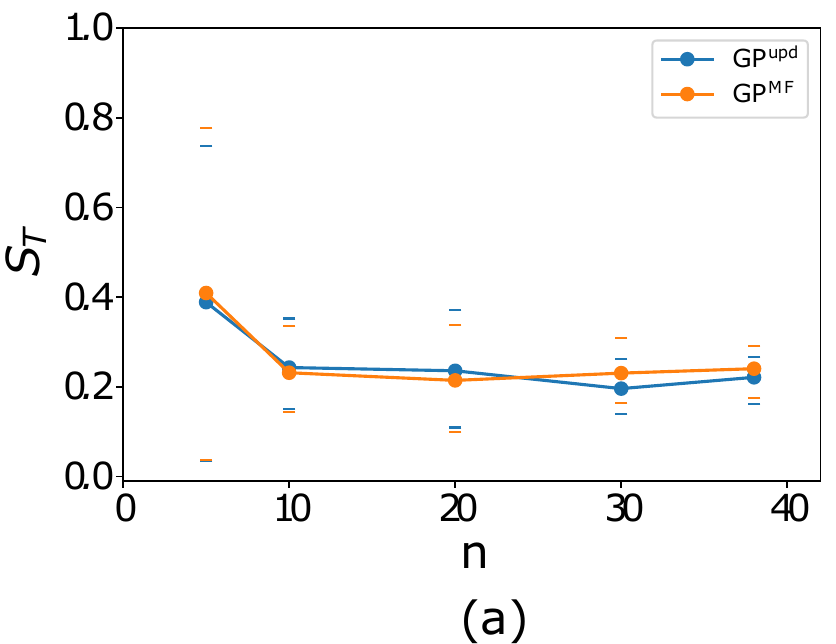} &
    \includegraphics[width=.49\textwidth]{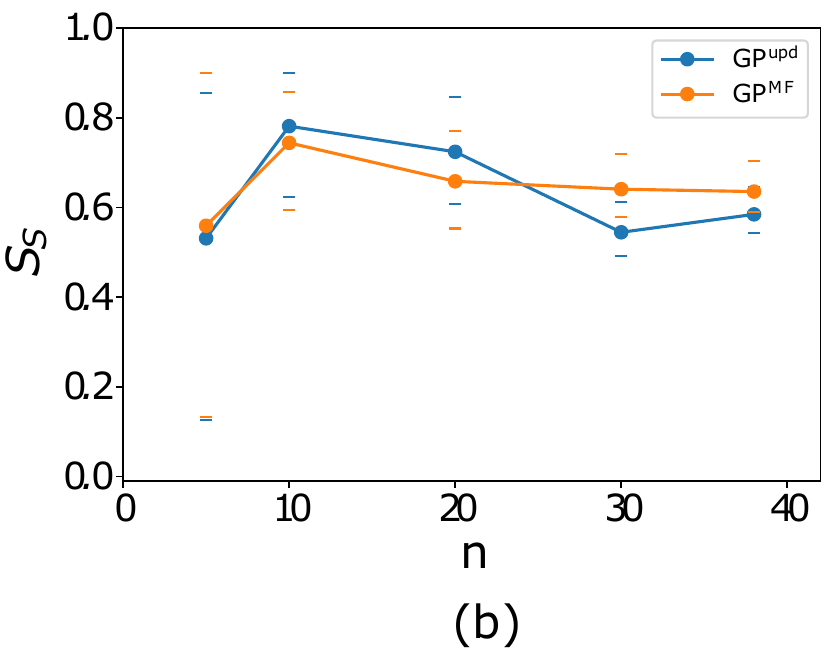}
  \end{tabular}
  \caption{First-order sensitivity index estimators for (a) nozzle temperature, and (b) nozzle speed, using two different pre-training and updating approaches for GP.}
  \label{fig:GP_am_first_pretrain}
\end{figure}

\begin{figure}[!htb]
\centering
  \begin{tabular}{@{}cc@{}}
    \includegraphics[width=.49\textwidth]{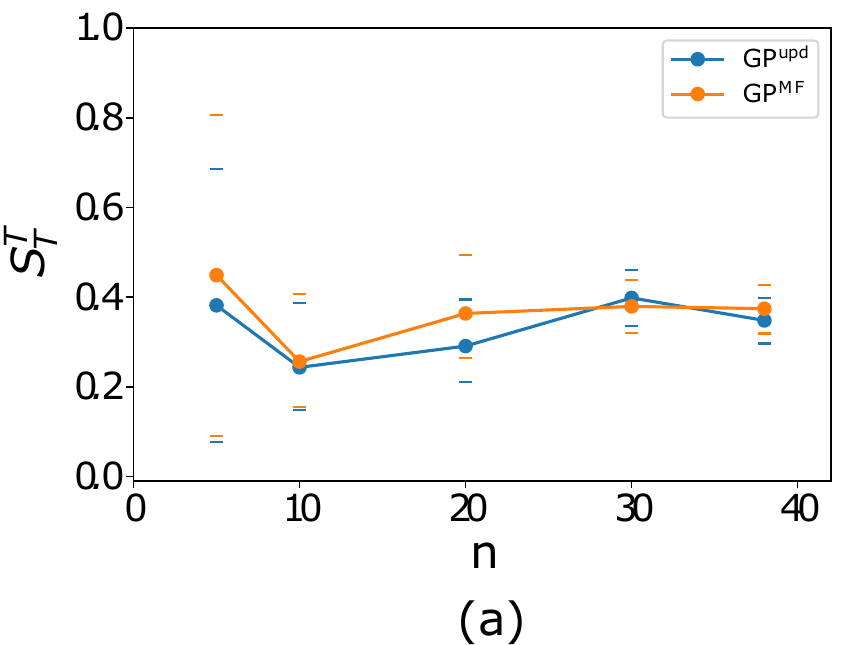} &
    \includegraphics[width=.49\textwidth]{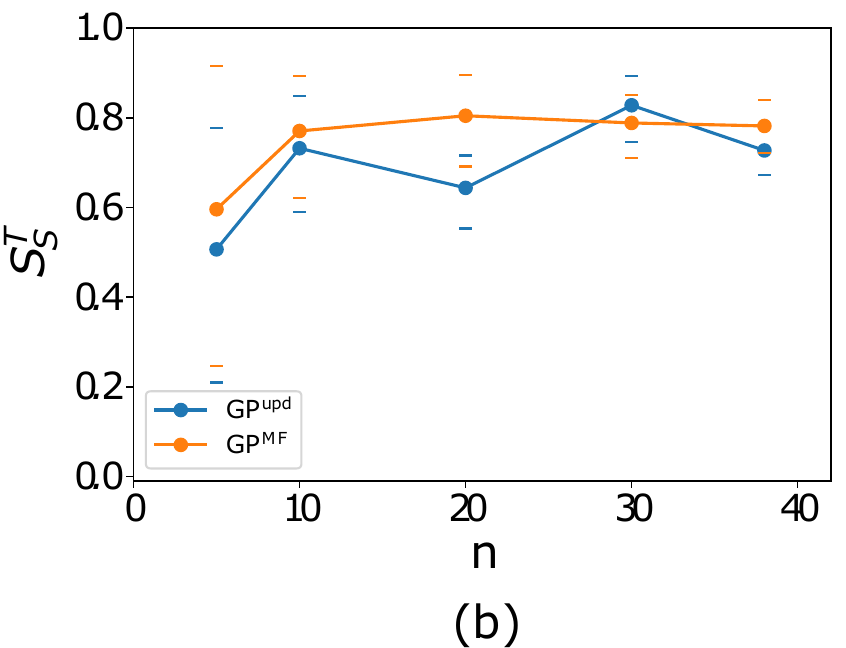}
  \end{tabular}
  \caption{Total effect sensitivity index estimates for (a) nozzle temperature, and (b) nozzle speed, using two different pre-training and updating approaches for GP.}
  \label{fig:GP_am_tot_pretrain}
\end{figure}

The Sobol' indices estimates obtained using the two approaches of the second PIML strategy are compared in Figs.~\ref{fig:GP_am_first_pretrain} and~\ref{fig:GP_am_tot_pretrain}. The results show that both approaches converge to similar first-order and total effect sensitivity index estimates. Thus, in the rest of the paper the first approach of the second PIML strategy is used for both GP and DNN models, i.e., $\mathbf{\rm GP^{\rm upd}}$ and $\mathbf{\rm DNN^{\rm upd}}$.

\subsubsection{GSA results using GP models}
\begin{figure}[!htb]
\centering
  \begin{tabular}{@{}cc@{}}
    \includegraphics[width=.49\textwidth]{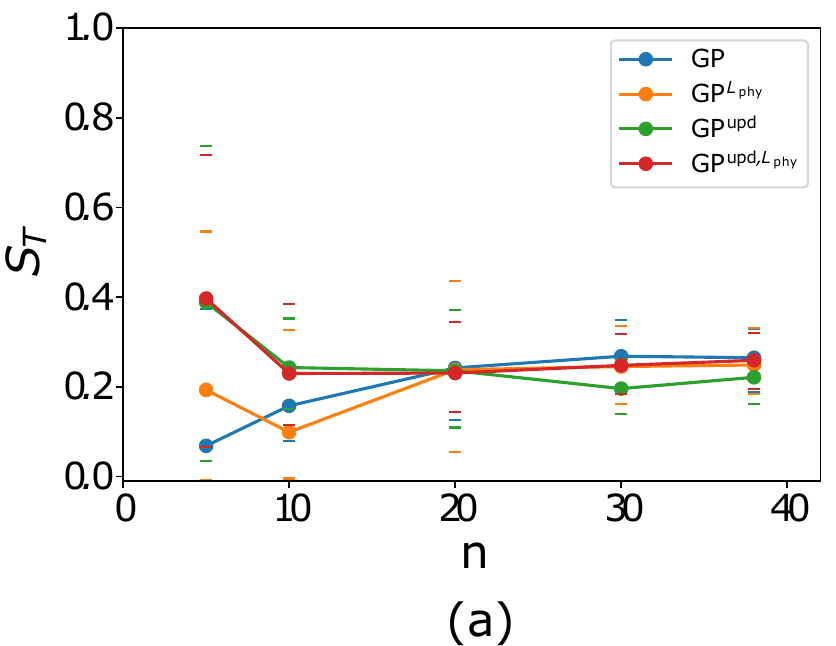} &
    \includegraphics[width=.49\textwidth]{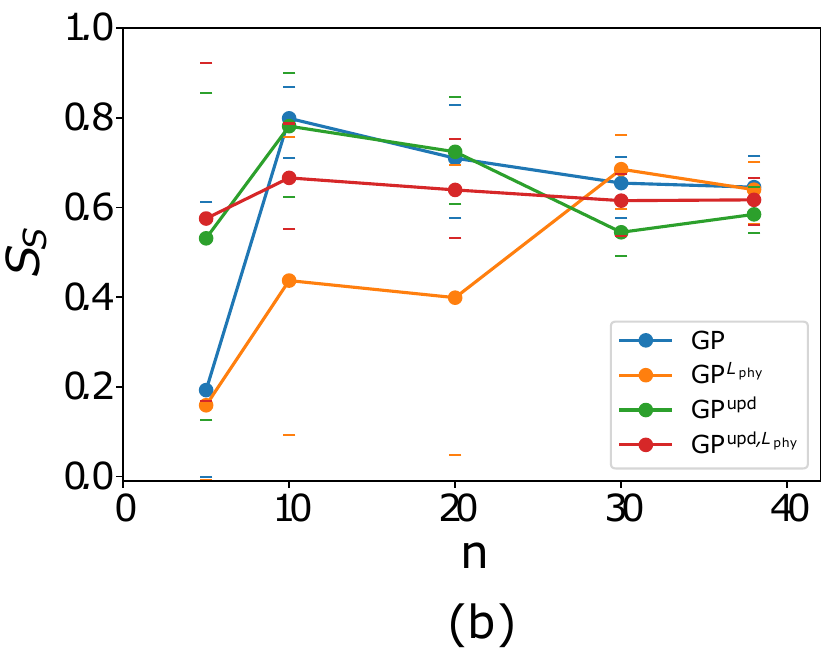}
  \end{tabular}
    \caption{First-order sensitivity index estimates for (a) nozzle temperature, and (b) nozzle speed, using the GP models.}
    \label{fig:gp_s1}
\end{figure}

\begin{figure}[!htb]
\centering
  \begin{tabular}{@{}cc@{}}
    \includegraphics[width=.49\textwidth]{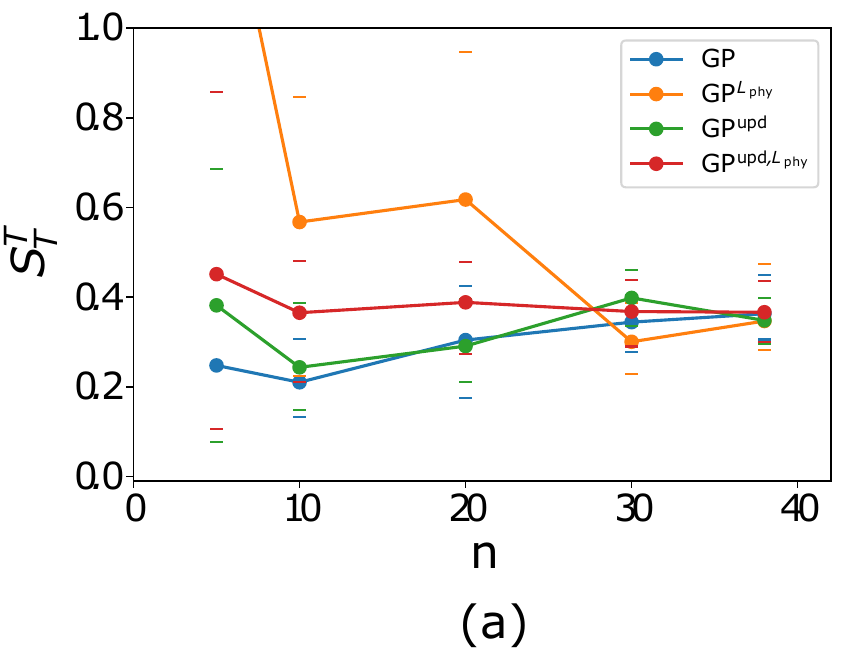} &
    \includegraphics[width=.49\textwidth]{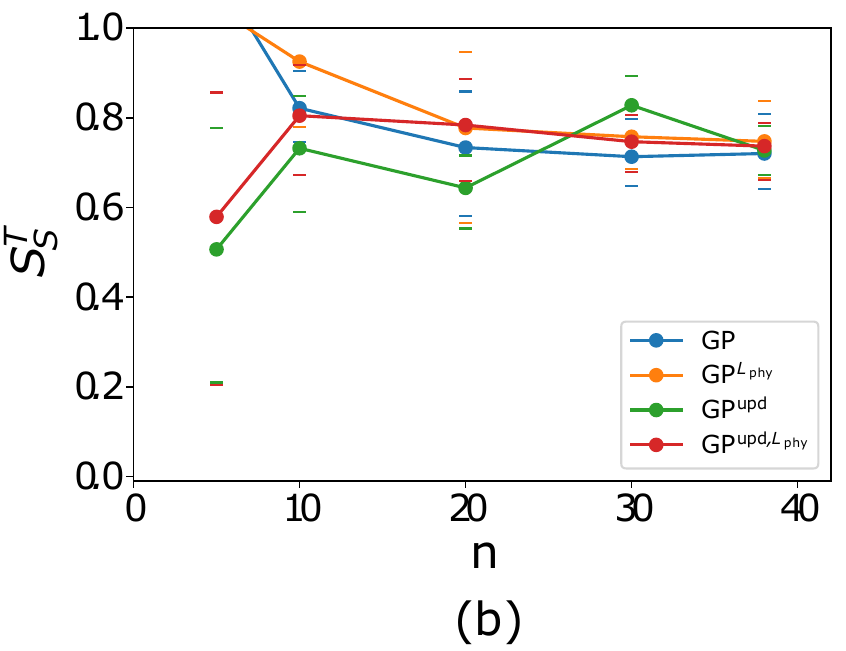}
  \end{tabular}
	\caption{Total effect sensitivity index estimates for (a) nozzle temperature, and (b) nozzle speed, using the GP models.}
	\label{fig:gp_sT}
\end{figure}

The Sobol' index computations with the GP models (1-4) are based on 5000 MC samples and 100 realizations of the Gaussian process. The effect of the number of experimental observations (used to train the GP models) on the first-order Sobol’ index estimates from the four GP models is illustrated in Fig.~\ref{fig:gp_s1}. And the total effect Sobol’ index estimates from the GP models for different numbers of experimental training data are shown in Fig.~\ref{fig:gp_sT}. The mean values of sensitivity estimates based on the GP model predictions are denoted with solid dots at a given number of observations $n$. The sensitivity results are reported for two input variables, nozzle temperature and nozzle speed; the output quantity of interest is porosity.

The 95\% prediction intervals are represented with bars above and below the solid dots for the corresponding model. The bounds in the sensitivity estimates are calculated using the mean predictions based on the 100 realizations of the GP models. As expected, the prediction intervals decrease as increasing amounts of experimental data are used to train the GP models. Further, all four models converge to similar first-order and total effect sensitivity estimates for both printer nozzle temperature and speed. The relative individual contribution (at $n$=39) of nozzle speed to the variance of the porosity ($\approx$ 0.65) is greater than that of the nozzle temperature ($\approx$ 0.25) and their sum is $\approx$ 0.9. And the sum of their total effect indices (which capture parameter interactions) is slightly above 1.0, indicating that the interaction effect is small.

For further clarity regarding uncertainty, numerical values of the prediction bounds of first-order sensitivity estimates of nozzle temperature obtained using 100 realizations of the GP models are shown in Table~\ref{table:Bounds}. The results indicate that prediction intervals obtained using the PIML models 3 and 4 ($\mathbf{\rm GP^{\rm upd}}$ and $\mathbf{\rm GP^{\rm upd, \mathcal{L}_{\rm phy}}}$) converge to the bounds obtained using 39 number of observations for training the models faster than the first two models. Note that the upper 95\% bound for Model 2 ($\mathbf{\rm GP^{\mathcal{L}_{\rm phy}}}$) converges to its final value (0.33) within 10 experimental observations. 
\begin{table*}[!htb]
\small
\setlength\tabcolsep{1.2pt}
    \caption{First-order sensitivity estimate prediction bounds of nozzle temperature for different amounts of experimental training data, using the GP models.}
    \label{table:Bounds}
\centering{%
    \begin{tabular*}{\linewidth}{@{\extracolsep{\fill}}l|*{5}c@{\extracolsep{\fill}} | *{5}c@{\extracolsep{\fill}}}
    \hline
            Models & \multicolumn{5}{c|}{Lower 95\% confidence limit} & \multicolumn{5}{c}{Upper 95\% confidence limit} \\
             & n=5 & n=10 & n=20 & n=30 & n=39 & n=5 & n=10 & n=20 & n=30 & n=39 \\
    \hline
            1. $\mathbf{\rm GP}$ & 0.00 & 0.08 & 0.13 & 0.20 & 0.19 & 0.37 & 0.25 & 0.34 & 0.35 & 0.33 \\
            2. $\mathbf{\rm GP^{\mathcal{L}_{\rm phy}}}$ & 0.00 & 0.00 & 0.05 & 0.16 & 0.18 & 0.55 & 0.33 & 0.43 & 0.34 & 0.33 \\
            3. $\mathbf{\rm GP^{\rm upd}}$ & 0.03 & 0.14 & 0.10 & 0.16 & 0.17 & 0.77 & 0.33 & 0.34 & 0.31 & 0.29 \\
            4. $\mathbf{\rm GP^{\rm upd, \mathcal{L}_{\rm phy}}}$ & 0.06 & 0.11 & 0.14 & 0.18 & 0.20 & 0.71 & 0.38 & 0.34 & 0.31 & 0.32 \\
    \hline
    \end{tabular*}}%
\end{table*}

\subsubsection{GSA results using DNN models with MC dropout}
\begin{figure}[!ht]
	\centering
	\includegraphics[width=1.\textwidth]{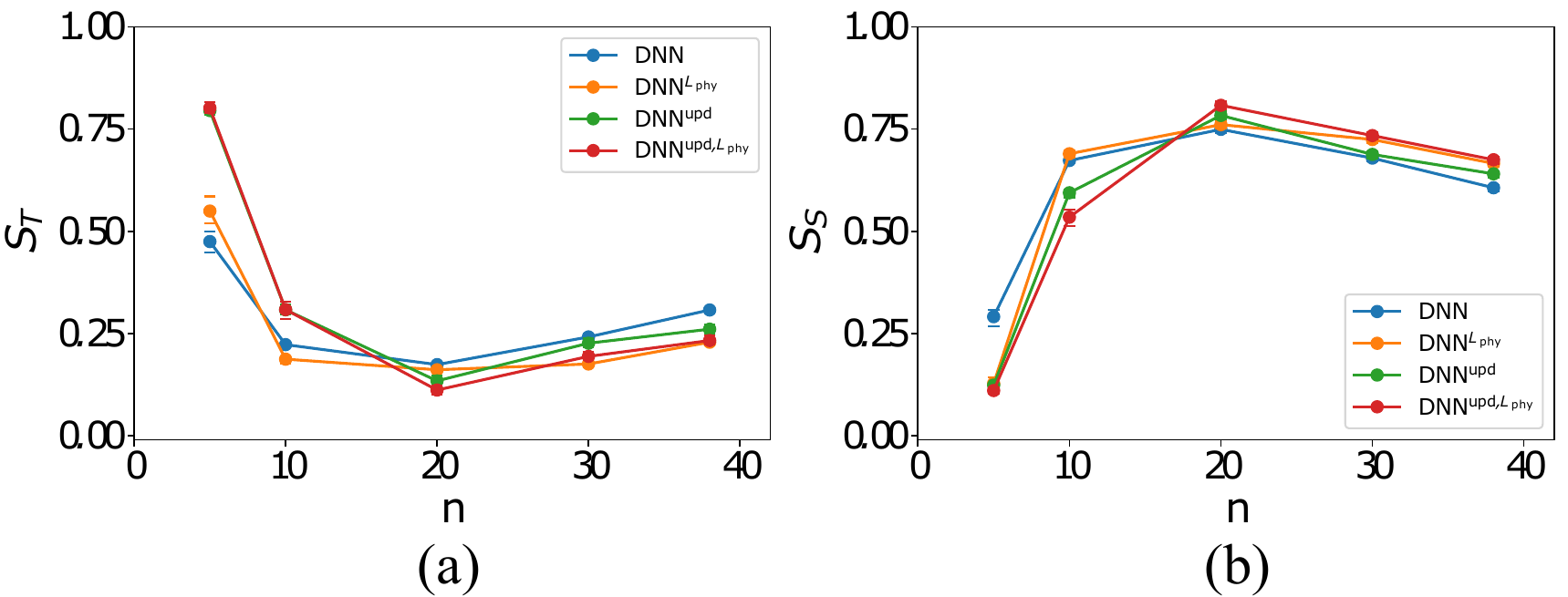}
	\caption{First-order sensitivity index estimators for (a) nozzle temperature, and (b) nozzle speed, using DNN models with MC dropout.}
	\label{fig:DNN_MC_s1}
\end{figure}

\begin{figure}[!ht]
	\centering
	\includegraphics[width=1.\textwidth]{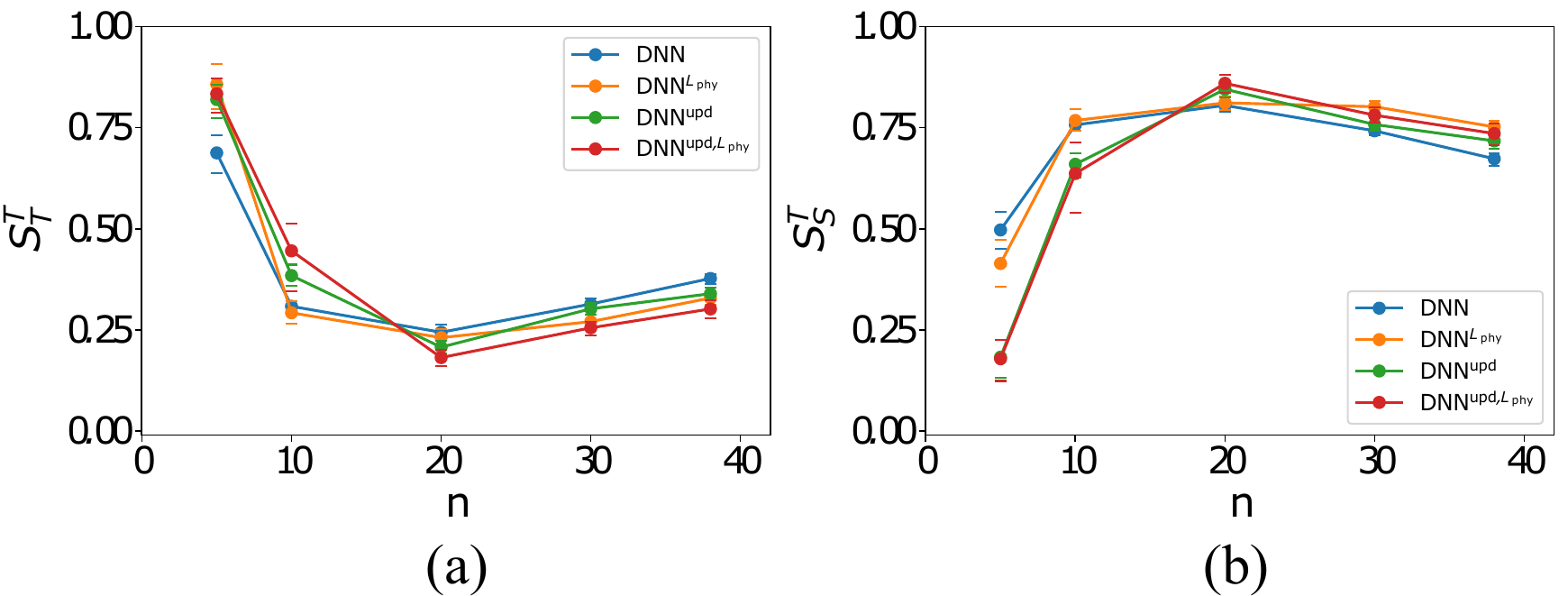}
	\caption{Total effect sensitivity index estimators for (a) nozzle temperature, and (b) nozzle speed, using DNN models with MC dropout.}
	\label{fig:DNN_MC_sT}
\end{figure}

The Sobol' index computations with the DNN models (5-8) are based on 5000 MC samples and 100 stochastic forward passes through the networks. The calculated first-order sensitivity estimates of temperature and speed, $S_T$ and $S_S$ respectively, for different numbers of experimental observations (training data) $n=(5,10,20,30,39 )$ are shown in Fig.~\ref{fig:DNN_MC_s1}. The distributions of sensitivity estimates are obtained using MC dropout predictions based on 100 stochastic forward passes through the networks for different number of observations. The mean values of sensitivity estimates are represented with solid dots and the 95\% bounds are denoted with bars above and below the solid dots for the corresponding model. Similarly, the calculated total effect sensitivity estimates $S_T$ and $S_S$ for different values of $n$ are illustrated in Fig.~\ref{fig:DNN_MC_sT}. Similar to the GP results, the difference between the total effect and first-order indices of process inputs is negligible, which indicates that their interaction is not significant.

All DNN models converge to similar first-order and total effect sensitivity estimates for both inputs, and these values are consistent with the results obtained using GP models. For this problem with two inputs and one output, 39 experimental observations appear adequate to train even the basic DNN model to achieve similar performance as the other physics-informed models; in fact, the results are similar even at 20 observations. However, the superior accuracy of the physics-informed models becomes apparent if only a small number of experiments are available, say $n=5$ or $n=10$, as shown in Table~\ref{table:EffectOfTrainingData_DNN} and discussed earlier in Section~\ref{sec:accuracy}.

For further clarity regarding uncertainty, numerical values of the prediction bounds of first-order sensitivity estimates of nozzle temperature obtained using 100 forward passes through the DNN models are given in Table~\ref{table:BoundsDNN}. The results show that prediction intervals obtained using all models show a similar trend. The 95\% bounds for all models are significantly narrower than the ones obtained using the GP models. The uncertainty in the sensitivity estimates due to the DNN models is almost negligible when more than 10 number of observations are used to train the models.
\begin{table*}[!htb]
\small
\setlength\tabcolsep{1.2pt}
    \caption{First-order sensitivity estimate prediction bounds of nozzle temperature for different amounts of experimental training data, using the DNN models.}
    \label{table:BoundsDNN}
\centering{%
    \begin{tabular*}{\linewidth}{@{\extracolsep{\fill}}l|*{5}c@{\extracolsep{\fill}} | *{5}c@{\extracolsep{\fill}}}
    \hline
            Models & \multicolumn{5}{c|}{Lower 95\% confidence limit} & \multicolumn{5}{c}{Upper 95\% confidence limit} \\
             & n=5 & n=10 & n=20 & n=30 & n=39 & n=5 & n=10 & n=20 & n=30 & n=39 \\
    \hline
            5. $\mathbf{\rm DNN}$ & 0.45 & 0.22 & 0.17 & 0.23 & 0.30 & 0.50 & 0.23 & 0.18 & 0.25 & 0.32 \\
            6. $\mathbf{\rm DNN^{\mathcal{L}_{\rm phy}}}$ & 0.52 & 0.18 & 0.15 & 0.17 & 0.22 & 0.59 & 0.19 & 0.17 & 0.18 & 0.24 \\
            7. $\mathbf{\rm DNN^{\rm upd}}$ & 0.78 & 0.30 & 0.12 & 0.22 & 0.25 & 0.81 & 0.32 & 0.15 & 0.24 & 0.27 \\
            8. $\mathbf{\rm DNN^{\rm upd, \mathcal{L}_{\rm phy}}}$ & 0.78 & 0.29 & 0.10 & 0.18 & 0.22 & 0.82 & 0.33 & 0.12 & 0.21 & 0.25 \\
    \hline
    \end{tabular*}}%
\end{table*}

The 95\% prediction intervals (upper limit-lower limit) decrease as increasing amounts of experimental data are used to train the DNN models. However, the prediction intervals obtained using the DNN models are much smaller than the ones obtained using the GP models since the DNN models has more degrees of freedom that can be optimized. For example, for $n=39$, the prediction interval width is 0.02-0.03 for the DNN models, whereas it is 0.12-0.15 for the GP models. In addition to the larger number of parameters, the number of training epochs, which is the number of complete passes through a batch of training dataset, is optimized for the DNN models, thus maximizing their prediction accuracy. The prediction accuracy is further improved by choosing the appropriate dropout rate as discussed earlier. The number of parameters to be learned reduces with the use of dropout, which helps with regularization and prevents ill-conditioning. Further, the  number of training epochs is also observed to affect both the accuracy and uncertainty of the sensitivity estimates. It was found that at a smaller number of epochs, the model was not fully trained resulting in underfitting, leading to larger bias and variance in the prediction. As the number of epochs was increased, both the bias and the variance were reduced.

\subsection{Illustrative example 2}
\subsubsection{Problem setup}
In order to further illustrate the effectiveness of the proposed method and how different combinations of methods affect the accuracy and computational effort, a more complex problem involving eleven random variables is considered. The data set used by Karpatne et al.~\cite{karpatne2017physics} for lake temperature modeling is considered, at Lake Mendota in Wisconsin, USA.

The overall data for Lake Mendota comprised of 13,543 temperature observations from 30 April 1980 to 02 Nov 2015. The modeling uses used 11 meteorological drivers as input variables. The original physics-based model, referred to as the General Lake Model (GLM), modeled the lake temperature by performing 1-D analysis (along depth) considering a variety of lake variables, and produced a total of 662,781 training samples (input-output). The physical non-linear relationship between temperature, depth and density of water is used to evaluate the physical violations across every consecutive depth-pair and time step. A more detailed description of the physics-based loss function and data can be found in the work by Karpatne et al.~\cite{karpatne2017physics}.

\subsubsection{Comparison of computational effort}
The computational costs of different models for training and estimation of Sobol’ indices based on 5000 MC samples with a fixed number of experimental data ($n=1000$) is given in Table.~\ref{table:Cost2}. The desktop computer used for the illustrative example 1 (Intel\textsuperscript{\tiny\textregistered} Xeon\textsuperscript{\tiny\textregistered} CPU E5-1660 v4$@$3.20GHz with 32 GB RAM and GPU NVIDIA Quadro K620 with 2 GB) is used for the computations of Sobol' indices. Among the GP models, the time it takes for training as well as computation of Sobol’ indices using Models 2-4 is similar to Model 1. The Sobol’ index estimations based on 5000 samples take approximately 1-3 minutes for the DNN models (Models 5-8). The training time and the computation of Sobol' indices using DNN models are significantly less than Models 1-4, which use the GP models, since the GP predictions require the covariance matrix to be inverted.
\begin{table*}[!htb]
    \caption{Computational effort of eight models for training and estimation of first-order and total effect Sobol' indices using 5000 MC samples with $n=1000$ number of temperature observations.}
    \label{table:Cost2}
\centering{%
    \begin{tabular*}{\linewidth}{@{\extracolsep{\fill}}l*{3}c@{\extracolsep{\fill}}}
    \toprule
            Models & Training [in minutes] & Sobol' indices calculation [in minutes] \\
    \midrule
            1. $\mathbf{\rm GP}$ & 37.5 & 40.1 \\
            2. $\mathbf{\rm GP^{\mathcal{L}_{\rm phy}}}$ & 37.9 & 41.2 \\
            3. $\mathbf{\rm GP^{\rm upd}}$ & 36.8 & 39.7 \\
            4. $\mathbf{\rm GP^{\rm upd, \mathcal{L}_{\rm phy}}}$ & 36.9 & 39.8 \\
            5. $\mathbf{\rm DNN}$ & 1.3 & 2.8  \\
            6. $\mathbf{\rm DNN^{\mathcal{L}_{\rm phy}}}$ & 2.7 & 5.7  \\
            7. $\mathbf{\rm DNN^{\rm upd}}$ & 1.4 & 2.8 \\
            8. $\mathbf{\rm DNN^{\rm upd, \mathcal{L}_{\rm phy}}}$ & 2.8 & 6.1  \\
    \bottomrule
    \end{tabular*}}%
\end{table*}

\subsubsection{GSA results using GP models}
\begin{figure}[!htb]
\centering
  \begin{tabular}{@{}ccc@{}}
    \includegraphics[width=.33\textwidth]{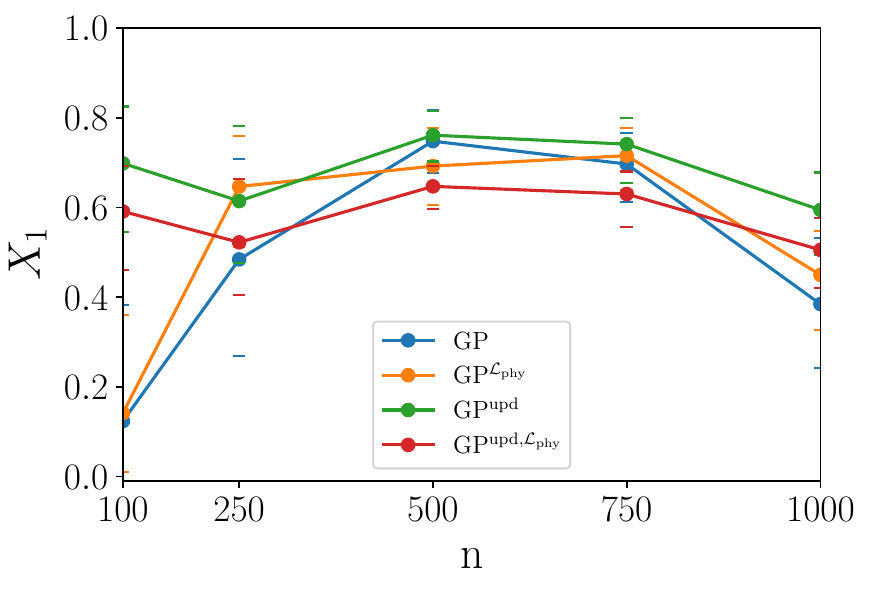} &
    \includegraphics[width=.33\textwidth]{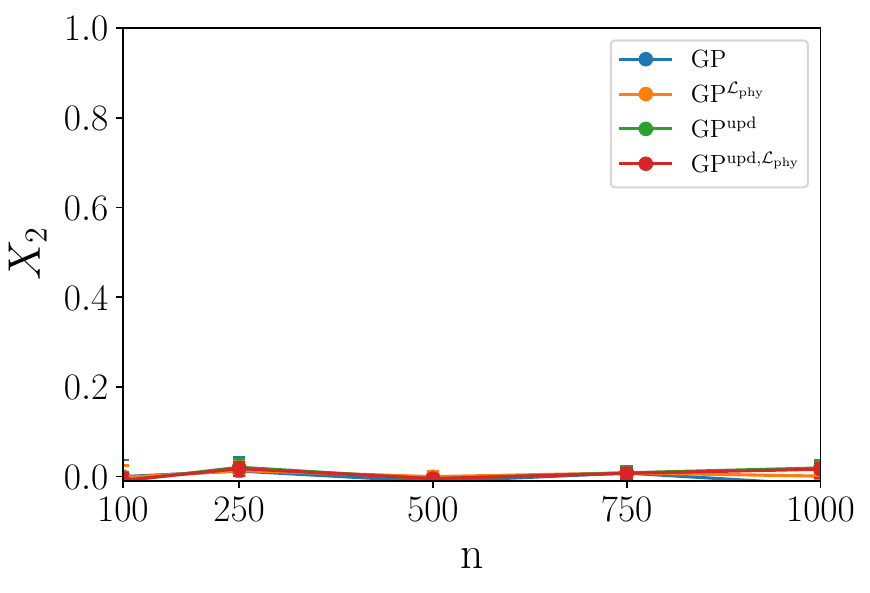} &
    \includegraphics[width=.33\textwidth]{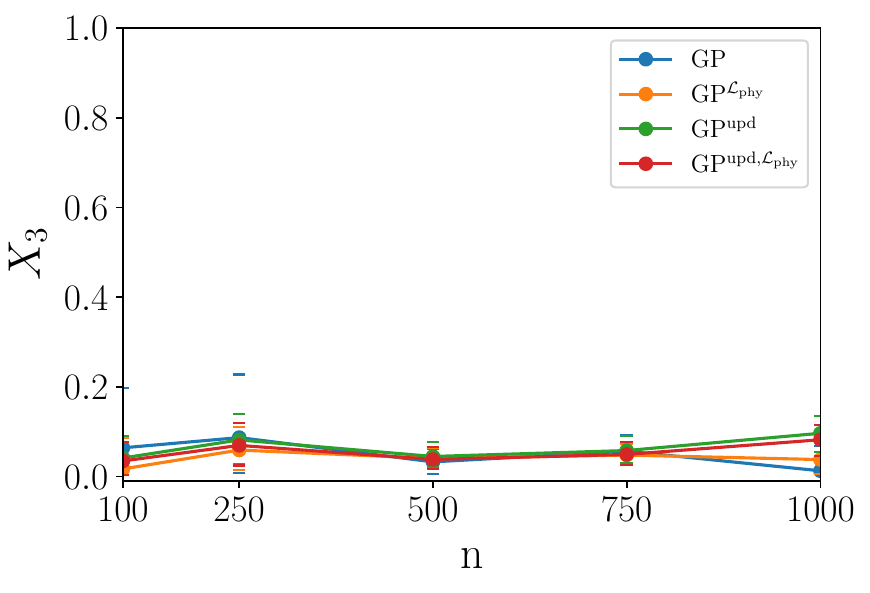} \\
    \includegraphics[width=.33\textwidth]{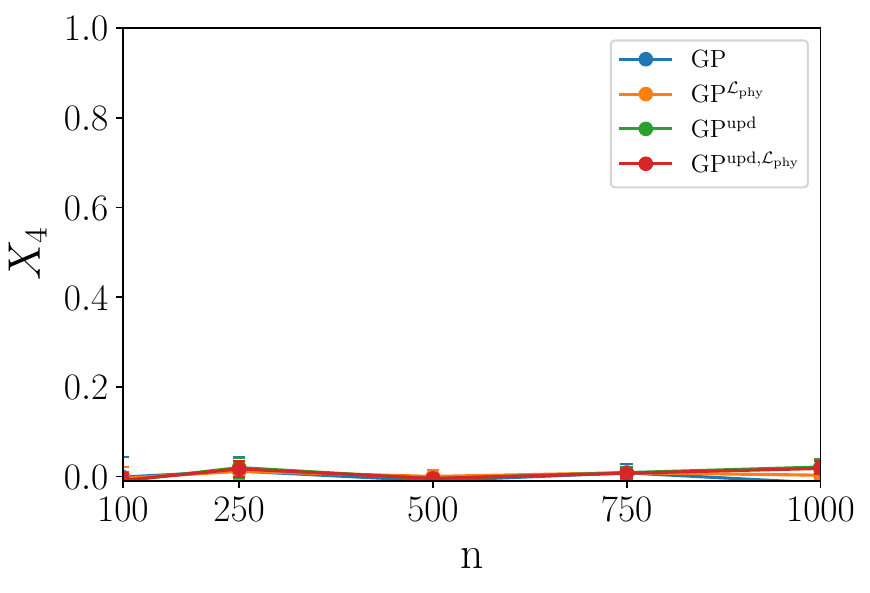} & 
    \includegraphics[width=.33\textwidth]{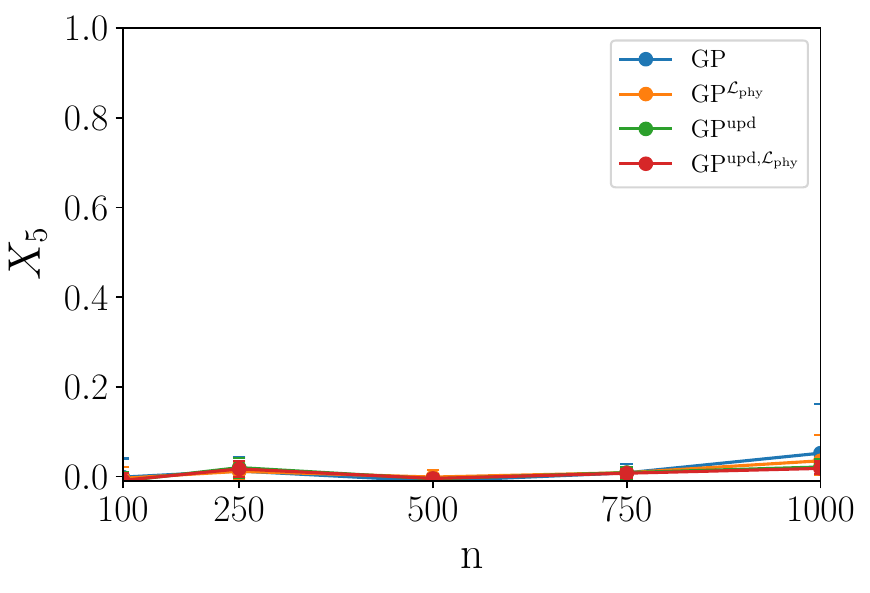} &
    \includegraphics[width=.33\textwidth]{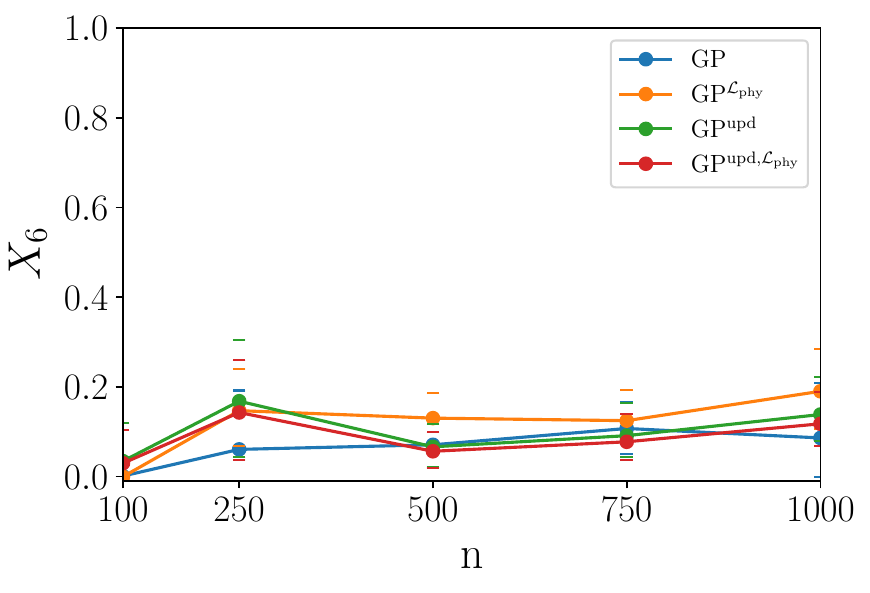} \\
    \includegraphics[width=.33\textwidth]{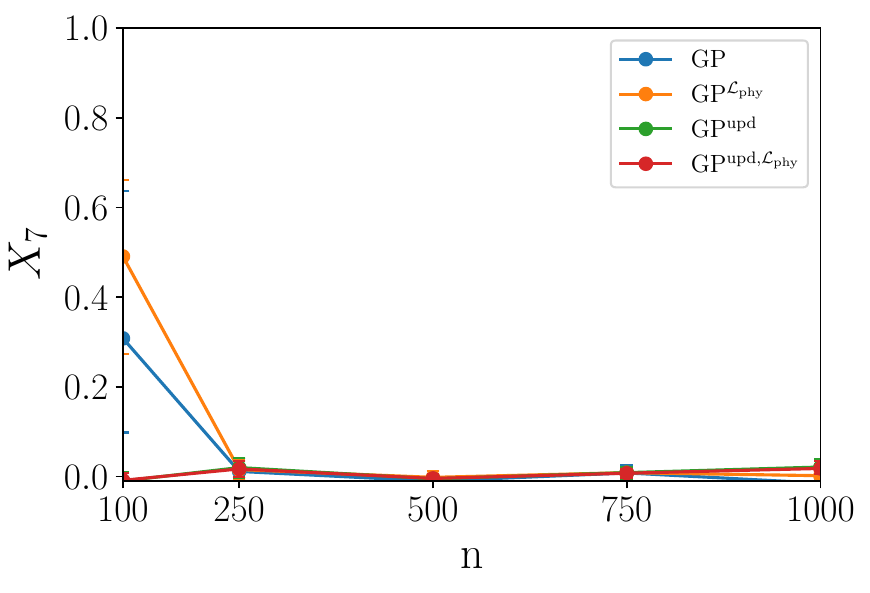} &
    \includegraphics[width=.33\textwidth]{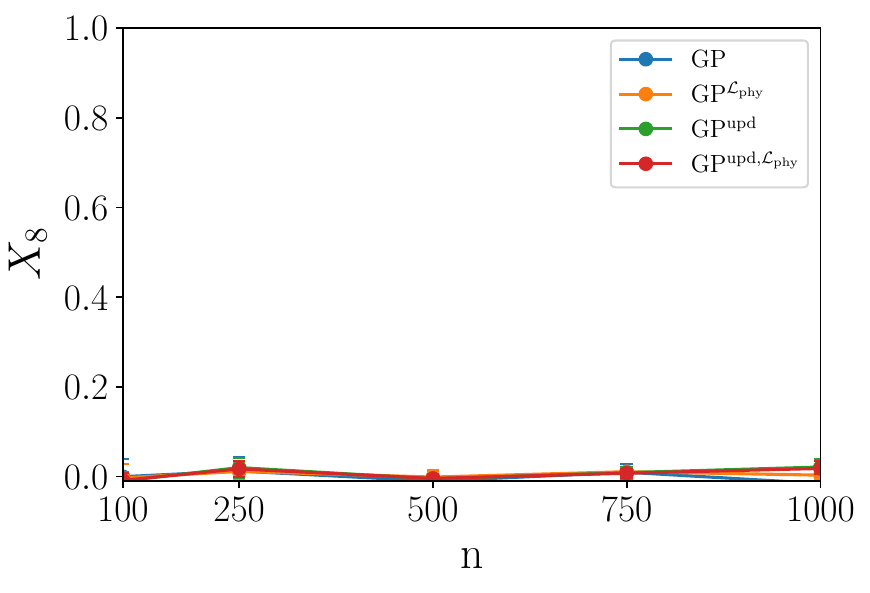} &
    \includegraphics[width=.33\textwidth]{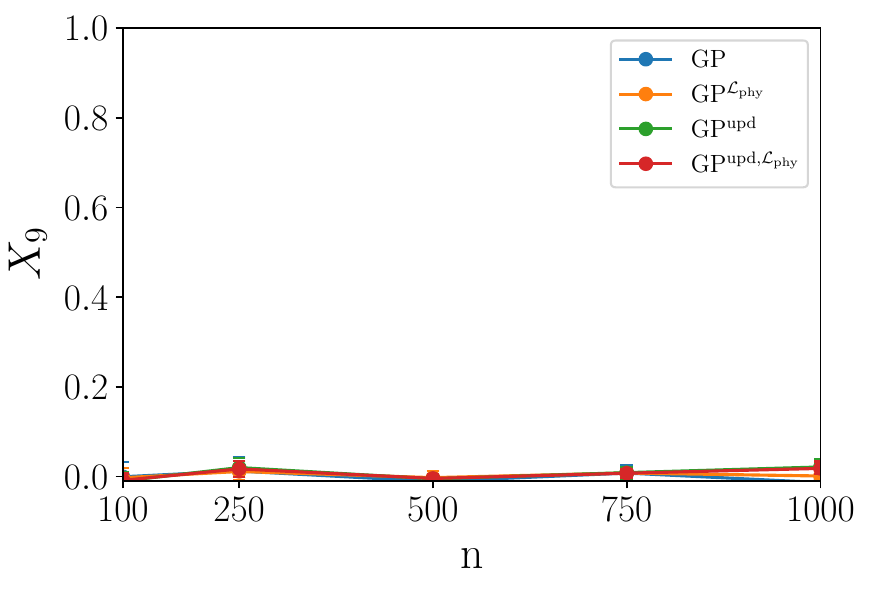} \\
    \includegraphics[width=.33\textwidth]{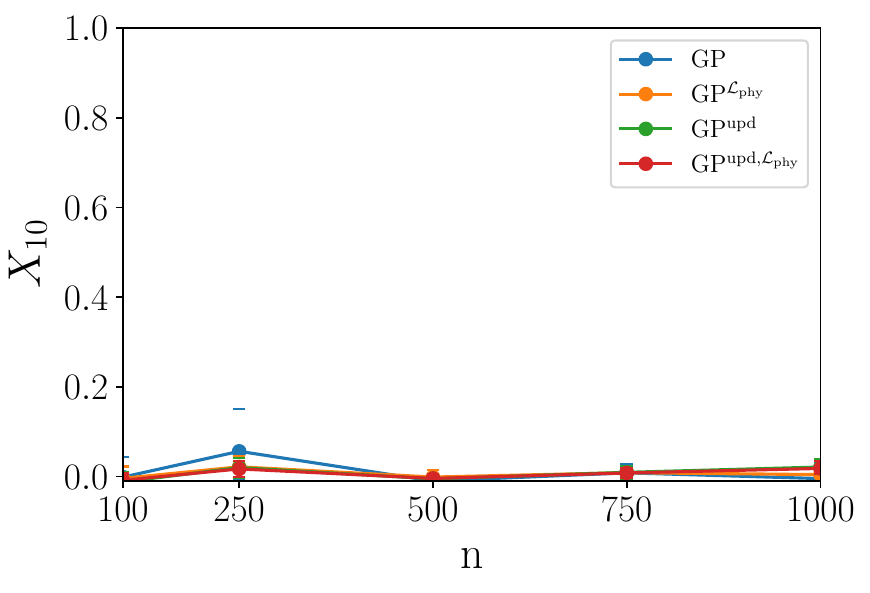} &
    \includegraphics[width=.33\textwidth]{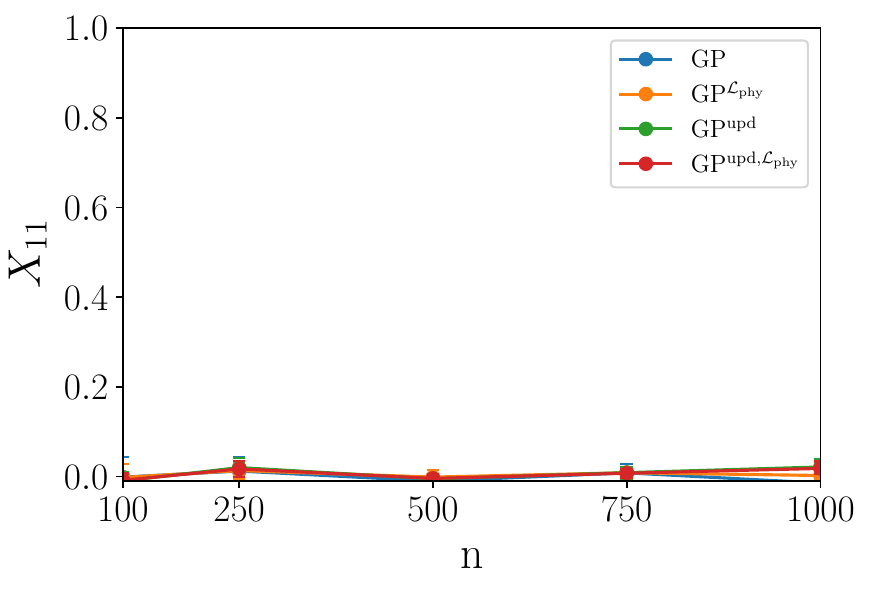} 
  \end{tabular}
  \caption{First-order sensitivity index estimators for the eleven variables of the lake temperature modeling using the GP models.}
  \label{fig:GP_lake_first}
\end{figure}

\begin{figure}[!htb]
\centering
  \begin{tabular}{@{}ccc@{}}
    \includegraphics[width=.33\textwidth]{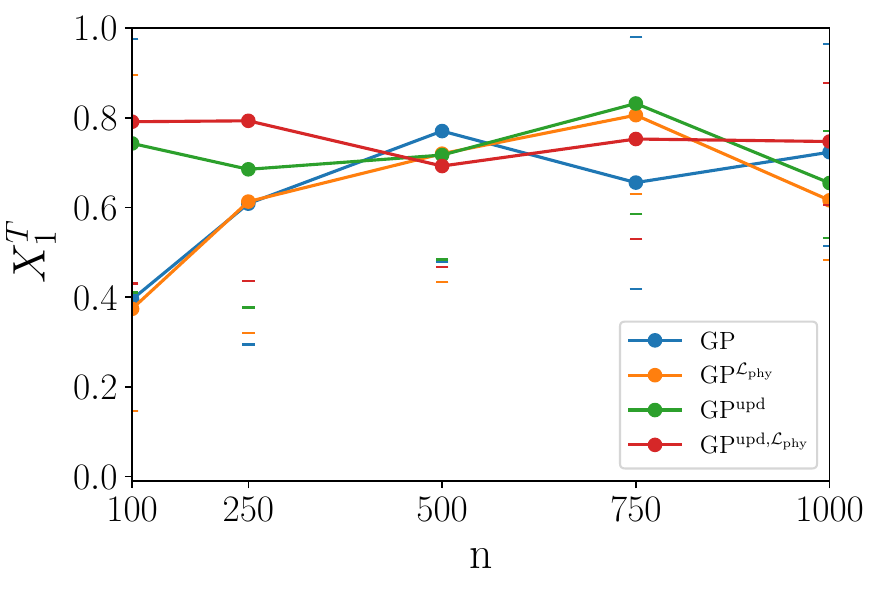} &
    \includegraphics[width=.33\textwidth]{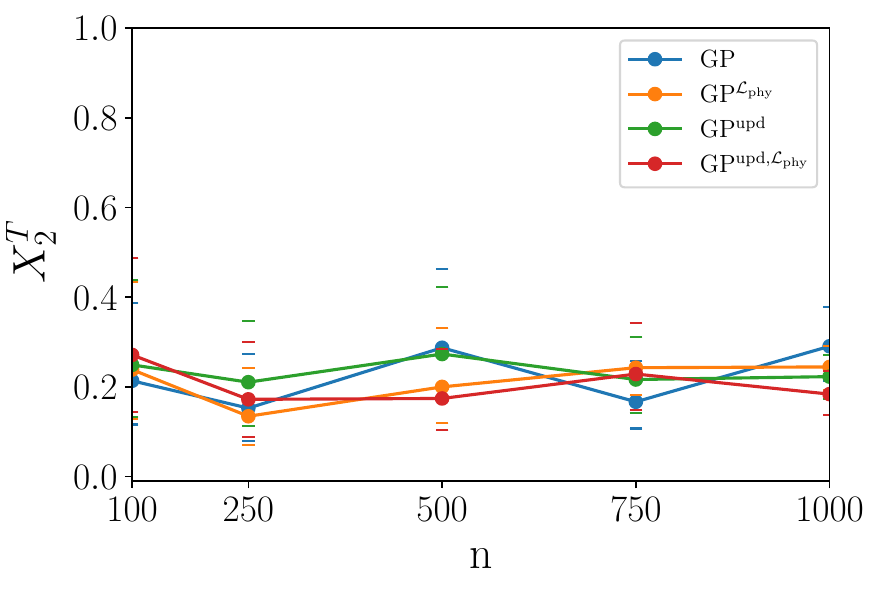} &
    \includegraphics[width=.33\textwidth]{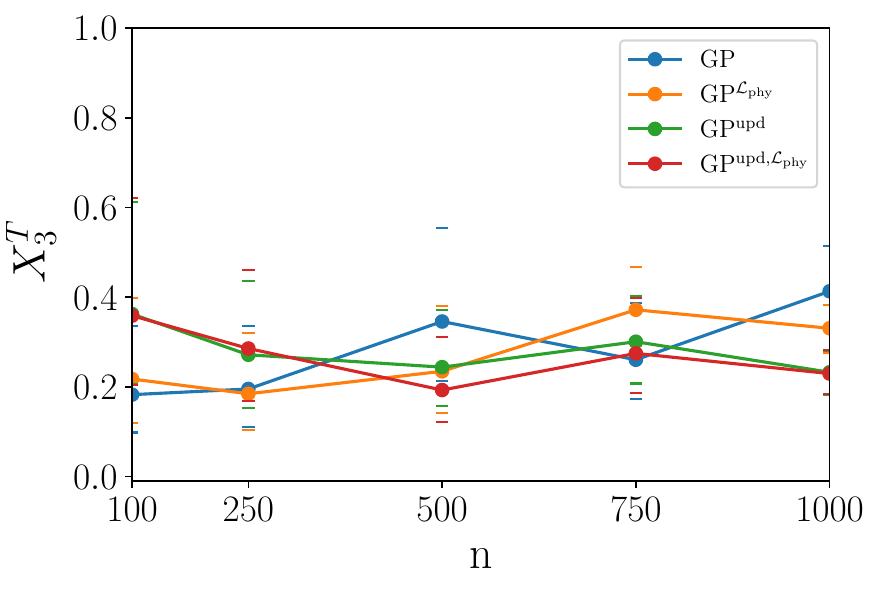} \\
    \includegraphics[width=.33\textwidth]{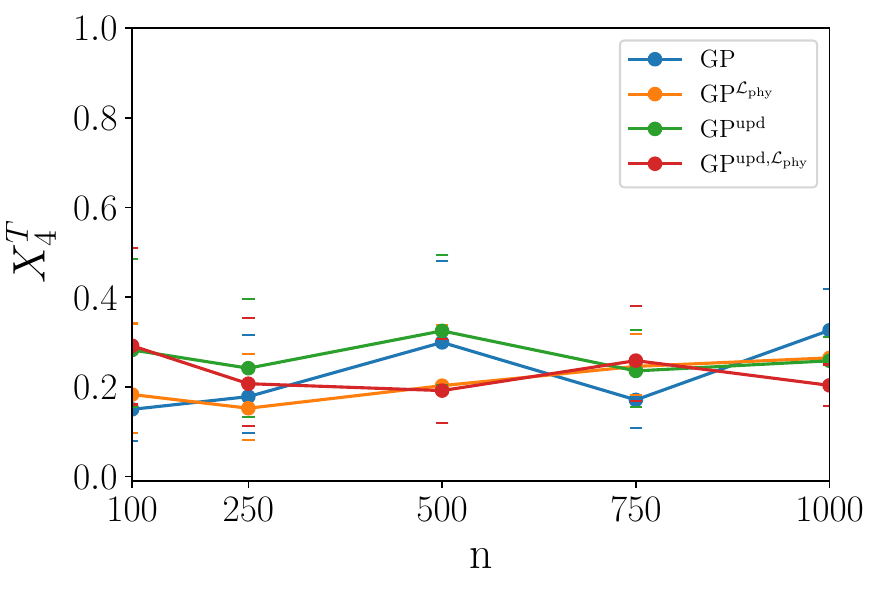} & 
    \includegraphics[width=.33\textwidth]{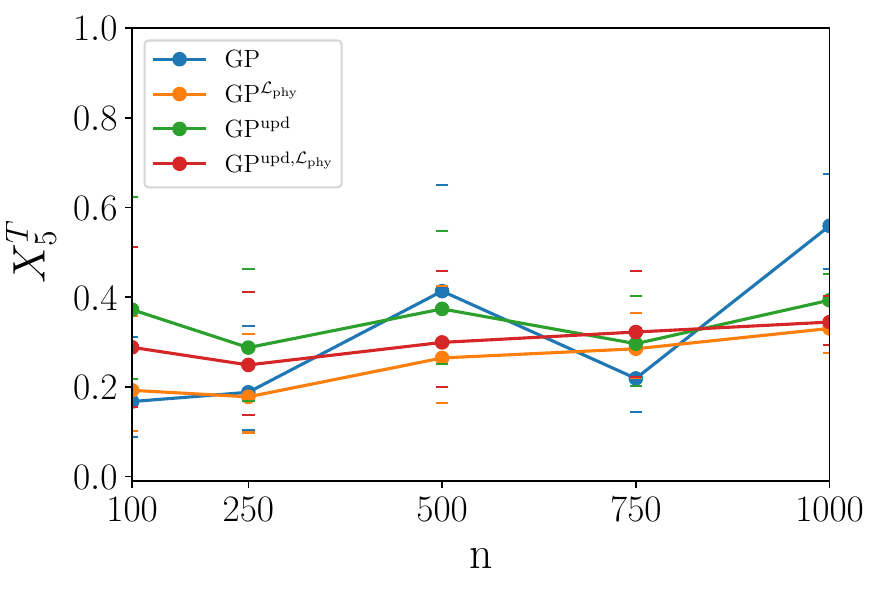} &
    \includegraphics[width=.33\textwidth]{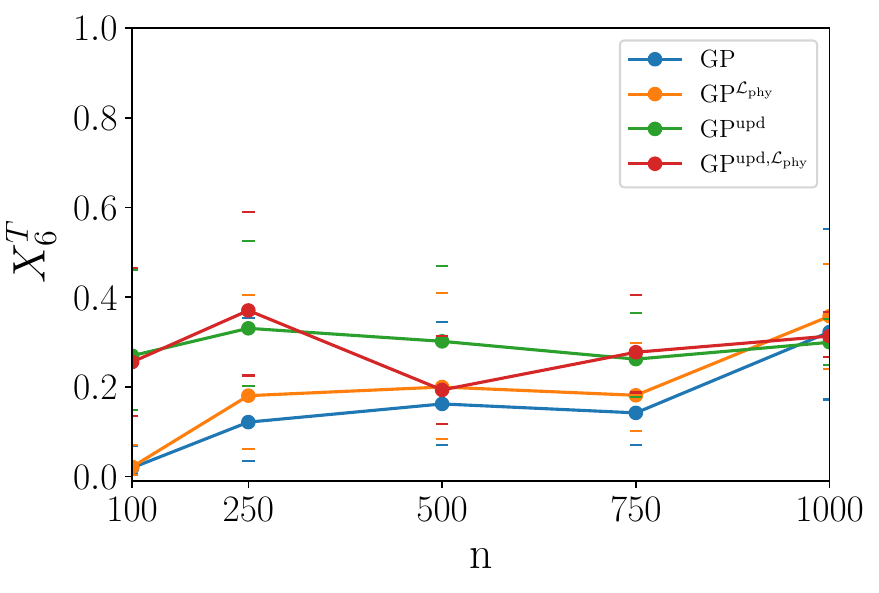} \\
    \includegraphics[width=.33\textwidth]{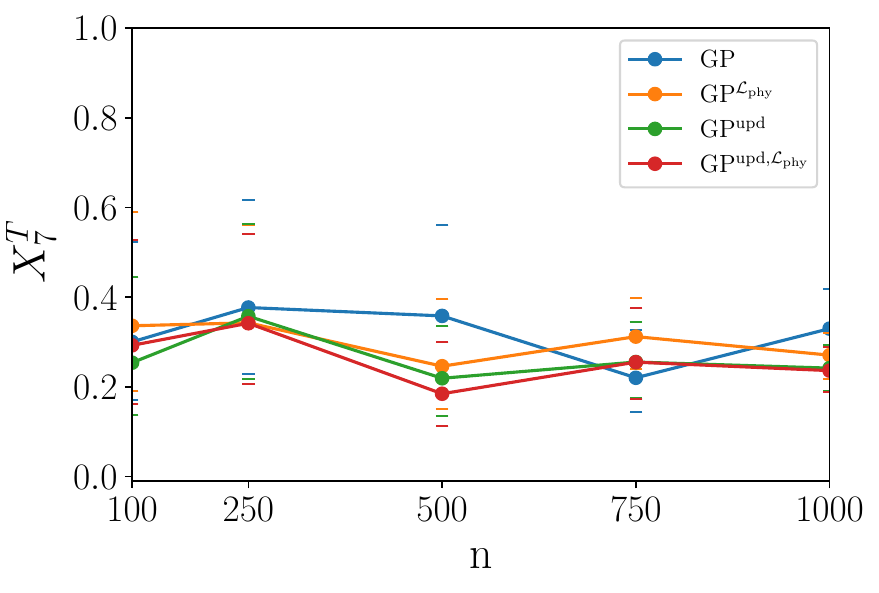} &
    \includegraphics[width=.33\textwidth]{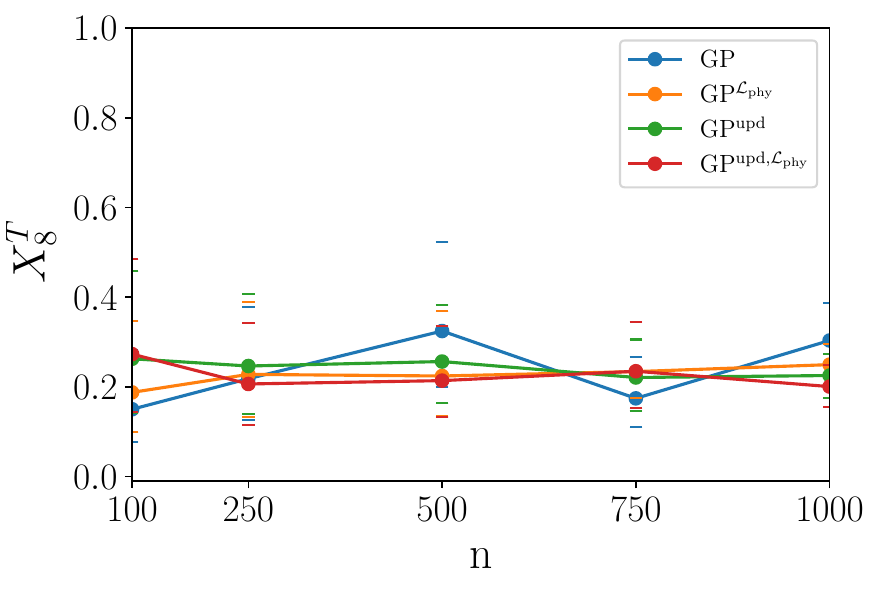} &
    \includegraphics[width=.33\textwidth]{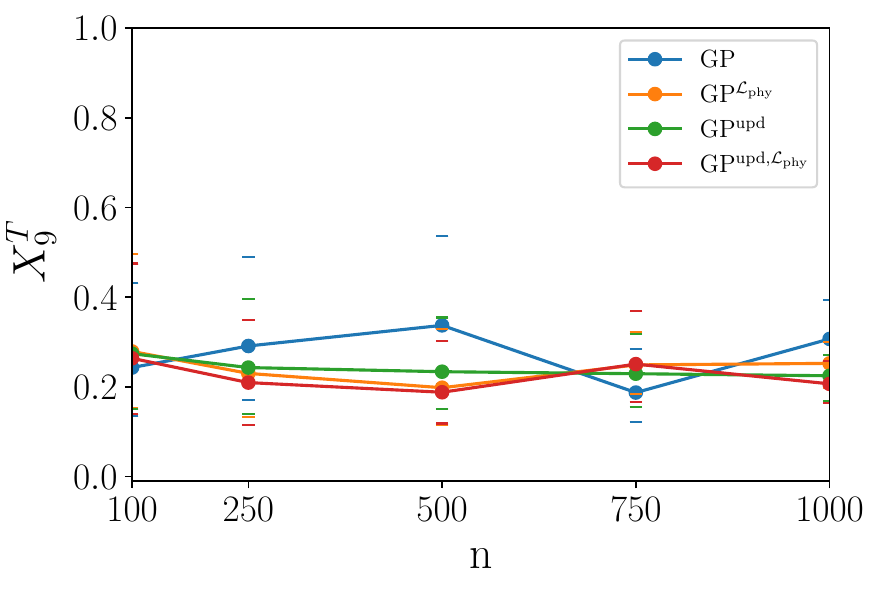} \\
    \includegraphics[width=.33\textwidth]{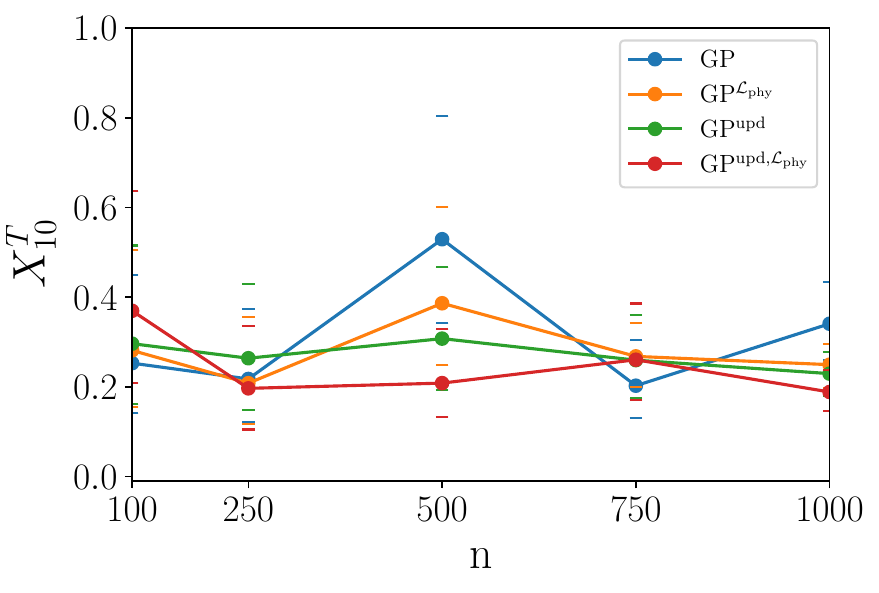} &
    \includegraphics[width=.33\textwidth]{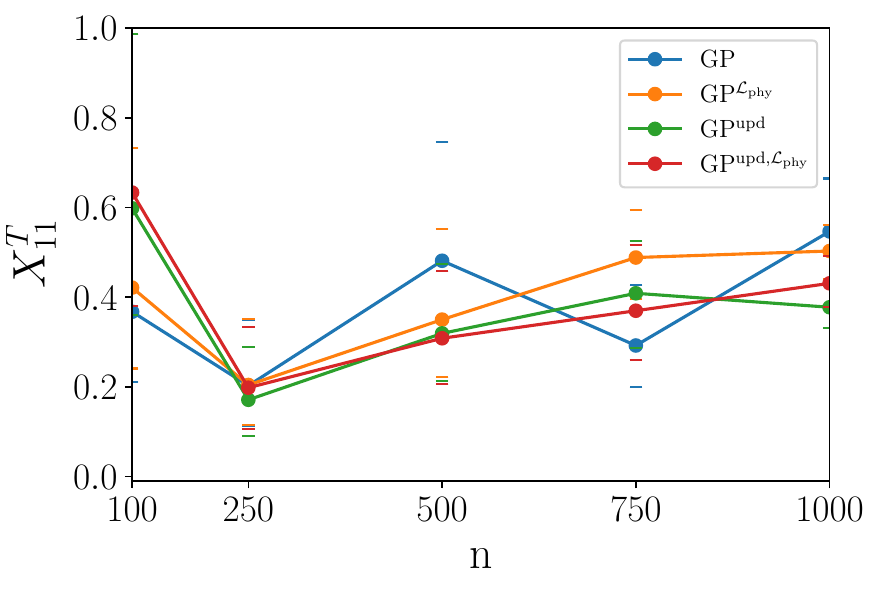} 
  \end{tabular}
  \caption{Total effect sensitivity index estimates for the eleven variables of the lake temperature modeling using the GP models.}
  \label{fig:GP_lake_tot}
\end{figure}

The effect of the number of experimental observations on the first-order and total effect Sobol’ index estimates from the four GP models of the lake temperature modeling is illustrated in Figs.~\ref{fig:GP_lake_first} and~\ref{fig:GP_lake_tot}.
The 95\% prediction intervals are represented with bars above and below the solid dots for the corresponding model. The prediction intervals decrease as increasing amounts of experimental data are used to train the GP models. In addition, all four models converge to similar first-order and total effect sensitivity estimates for all input variables. 

\subsubsection{GSA results using DNN models with MC dropout}
\begin{figure}[!htb]
\centering
  \begin{tabular}{@{}ccc@{}}
    \includegraphics[width=.33\textwidth]{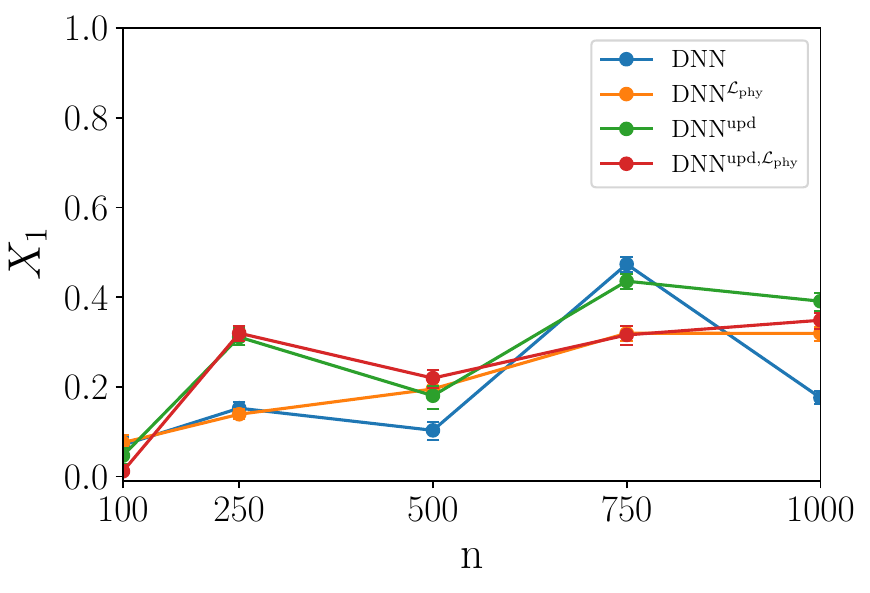} &
    \includegraphics[width=.33\textwidth]{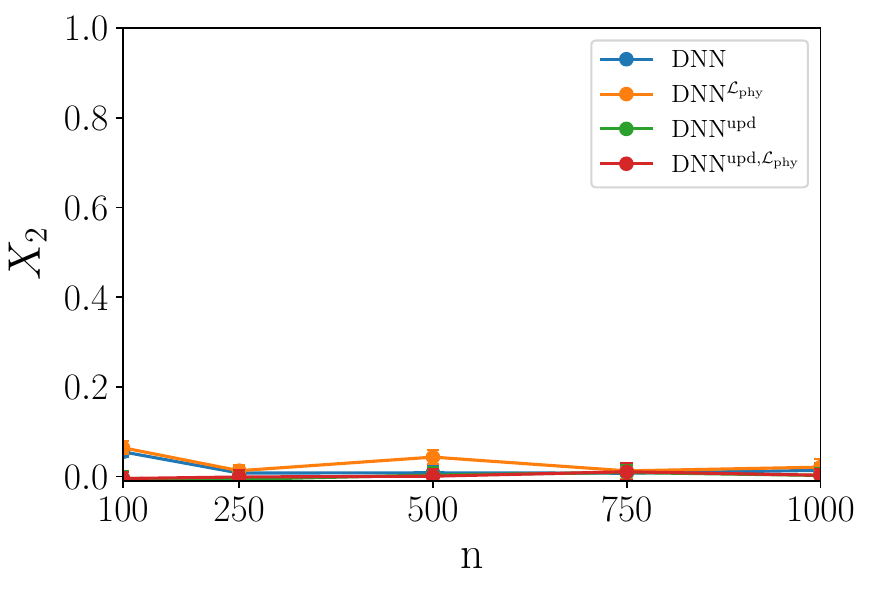} &
    \includegraphics[width=.33\textwidth]{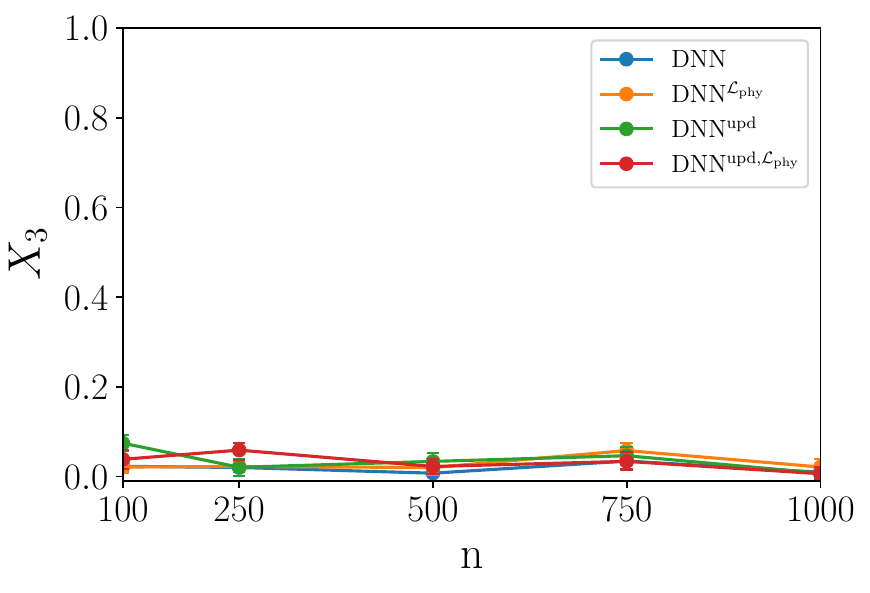} \\
    \includegraphics[width=.33\textwidth]{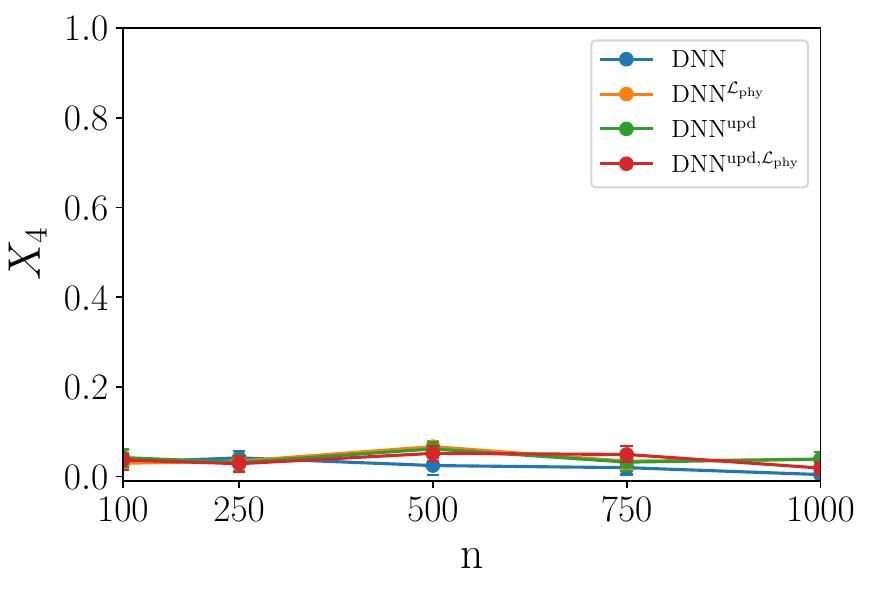} & 
    \includegraphics[width=.33\textwidth]{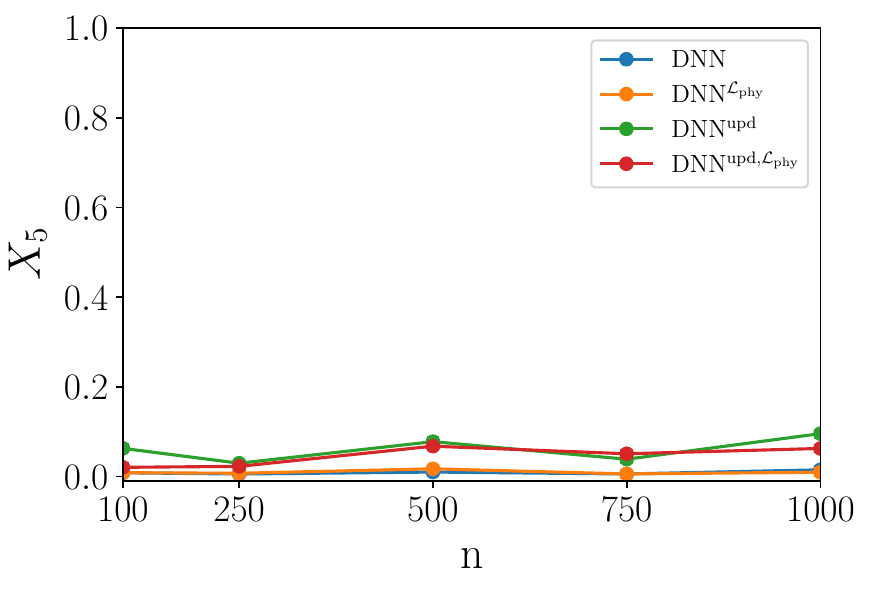} &
    \includegraphics[width=.33\textwidth]{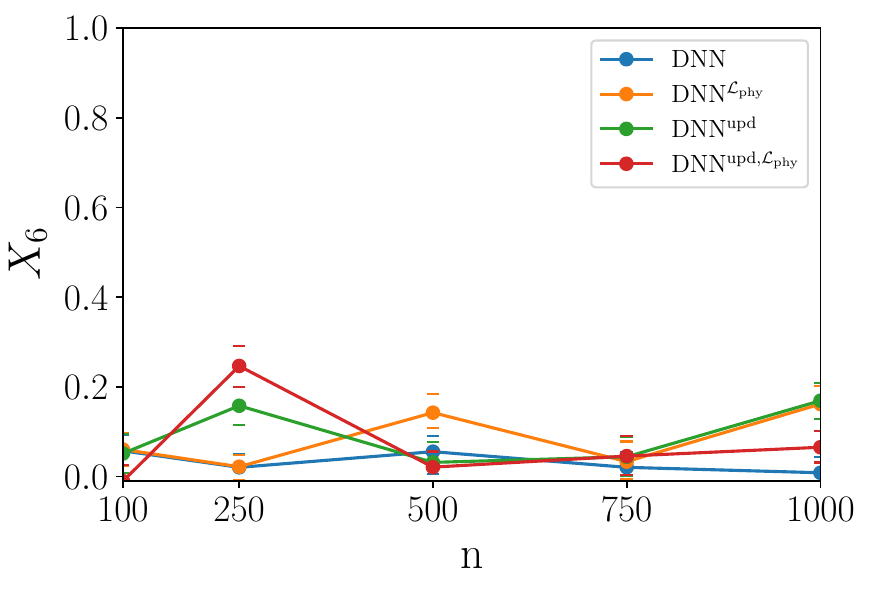} \\
    \includegraphics[width=.33\textwidth]{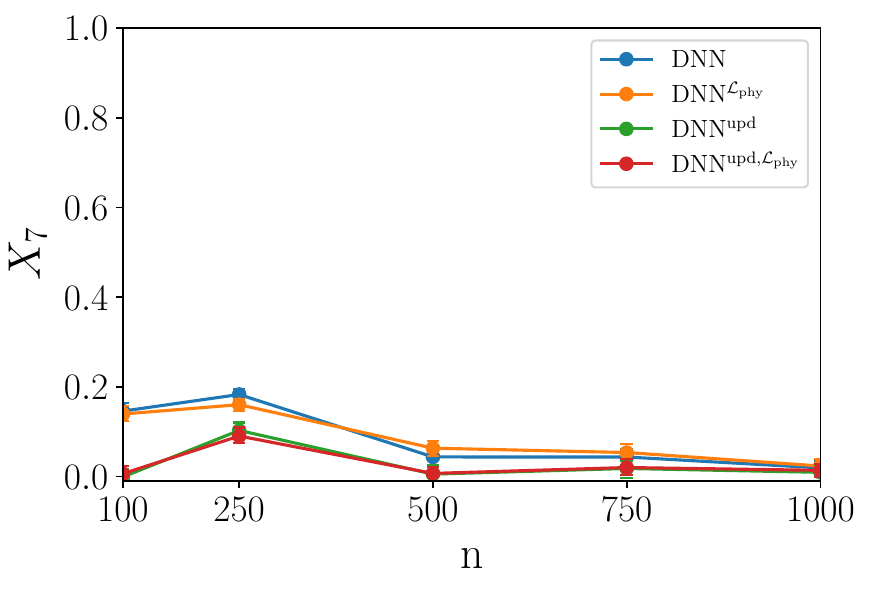} &
    \includegraphics[width=.33\textwidth]{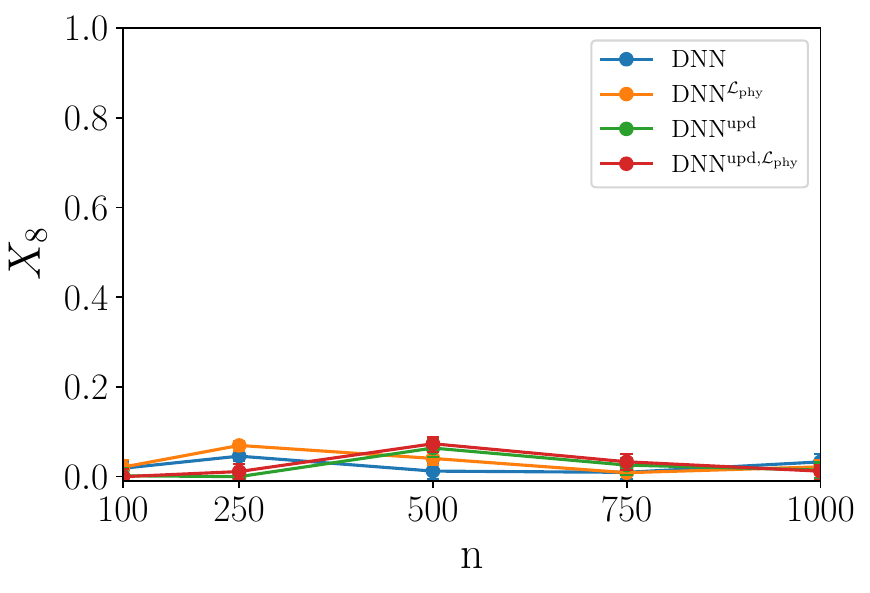} &
    \includegraphics[width=.33\textwidth]{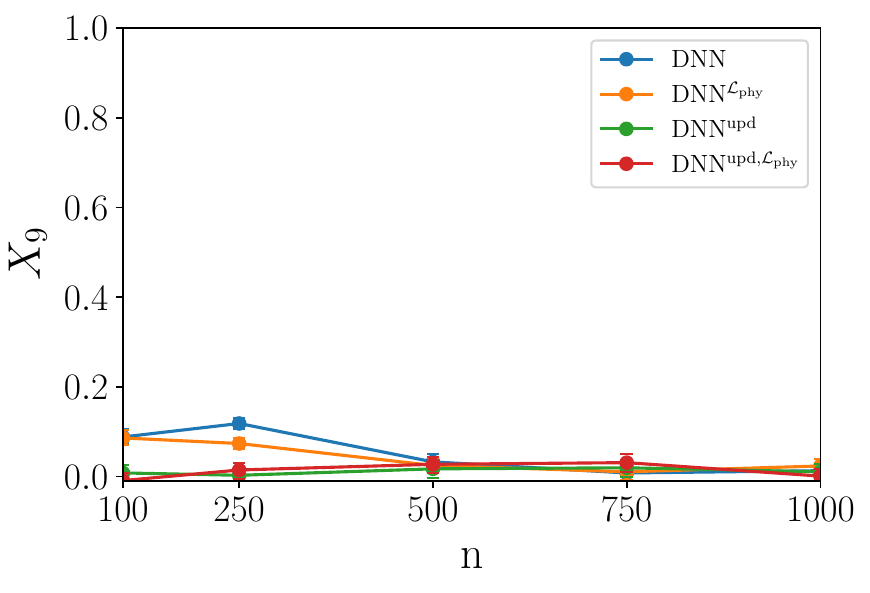} \\
    \includegraphics[width=.33\textwidth]{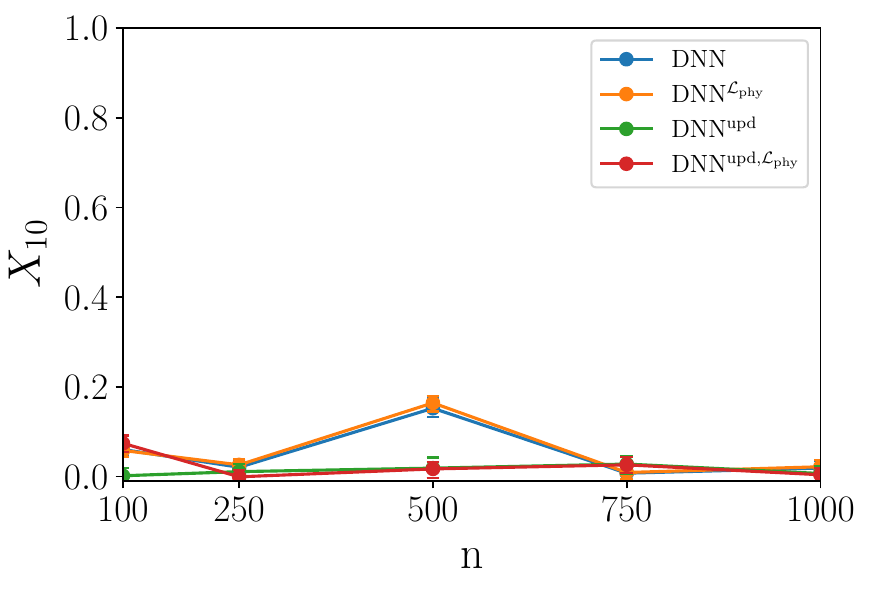} &
    \includegraphics[width=.33\textwidth]{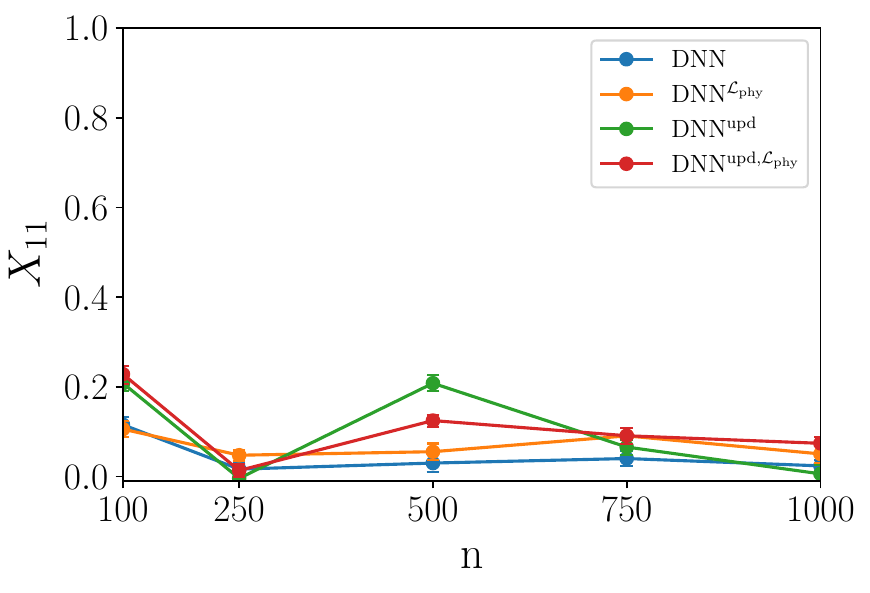} 
  \end{tabular}
  \caption{First-order sensitivity index estimators for the eleven variables of the lake temperature modeling using the DNN models.}
  \label{fig:DNN_lake_first}
\end{figure}

\begin{figure}[!htb]
\centering
  \begin{tabular}{@{}ccc@{}}
    \includegraphics[width=.33\textwidth]{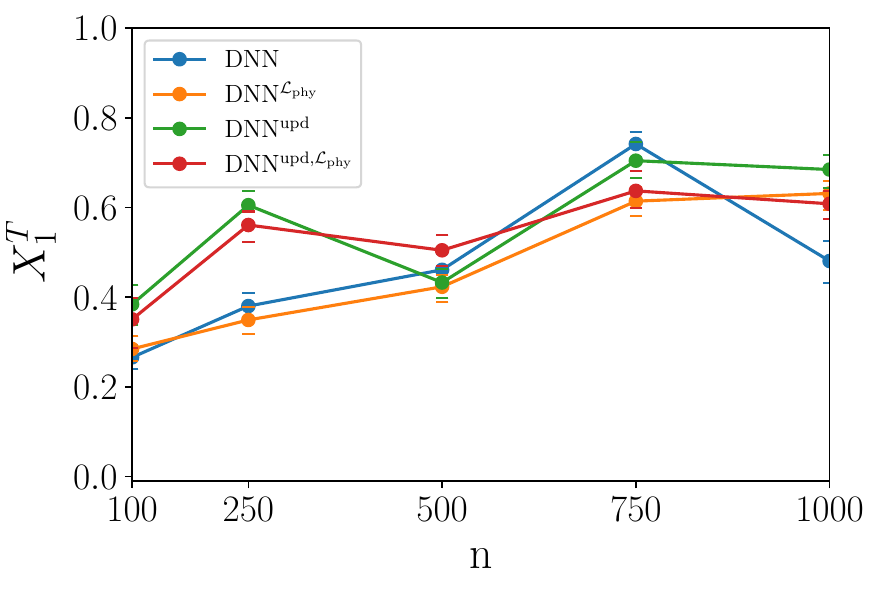} &
    \includegraphics[width=.33\textwidth]{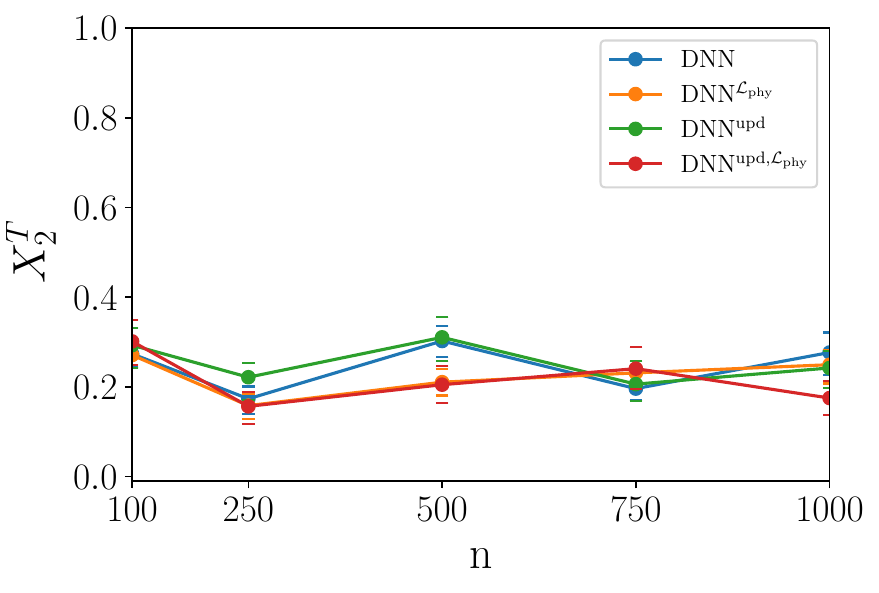} &
    \includegraphics[width=.33\textwidth]{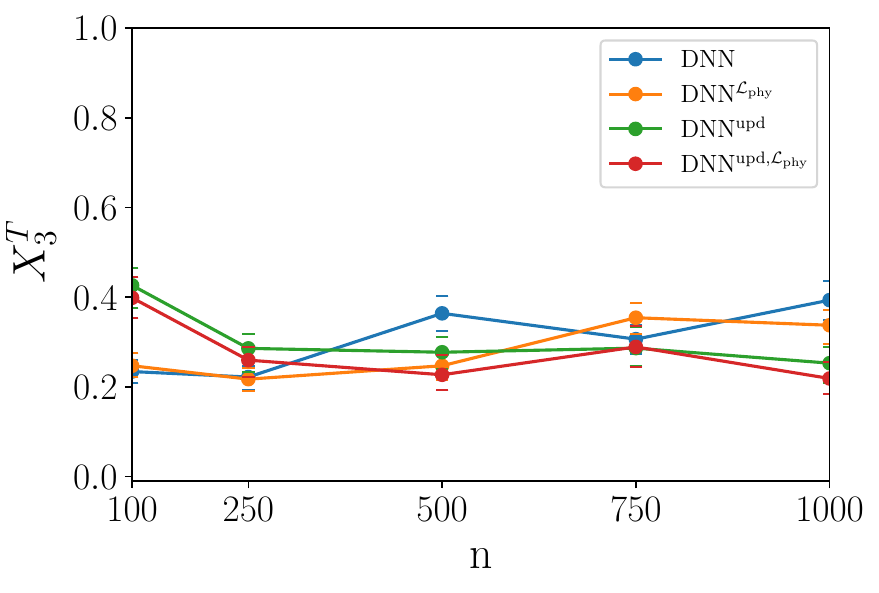} \\
    \includegraphics[width=.33\textwidth]{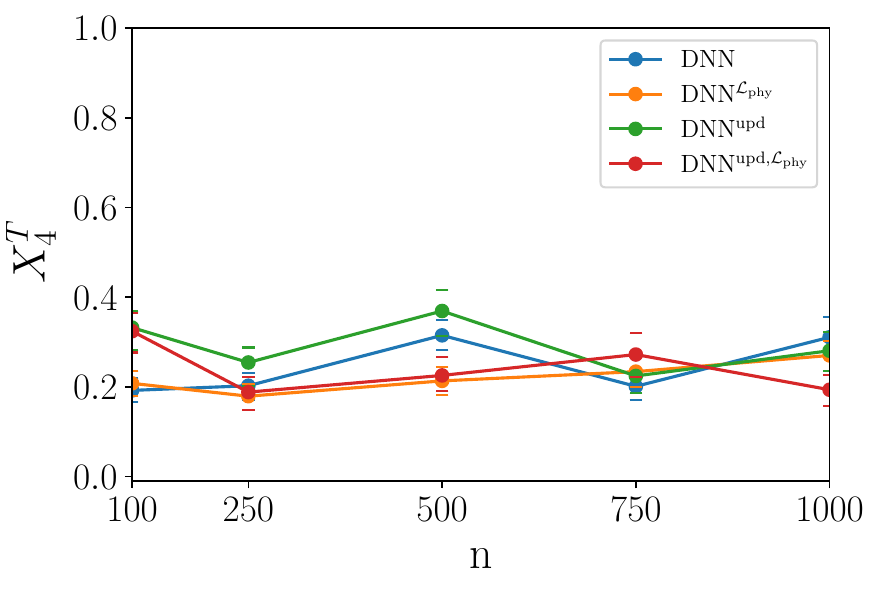} & 
    \includegraphics[width=.33\textwidth]{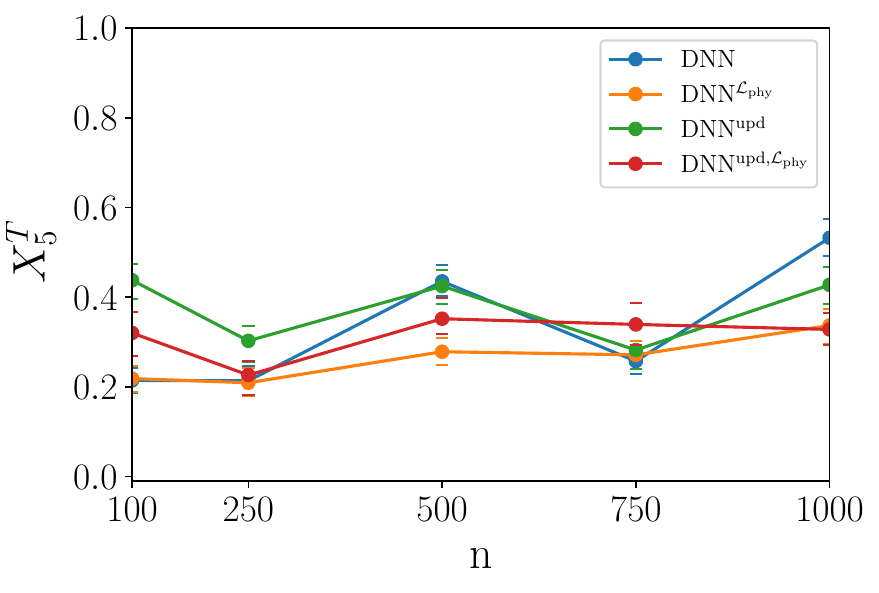} &
    \includegraphics[width=.33\textwidth]{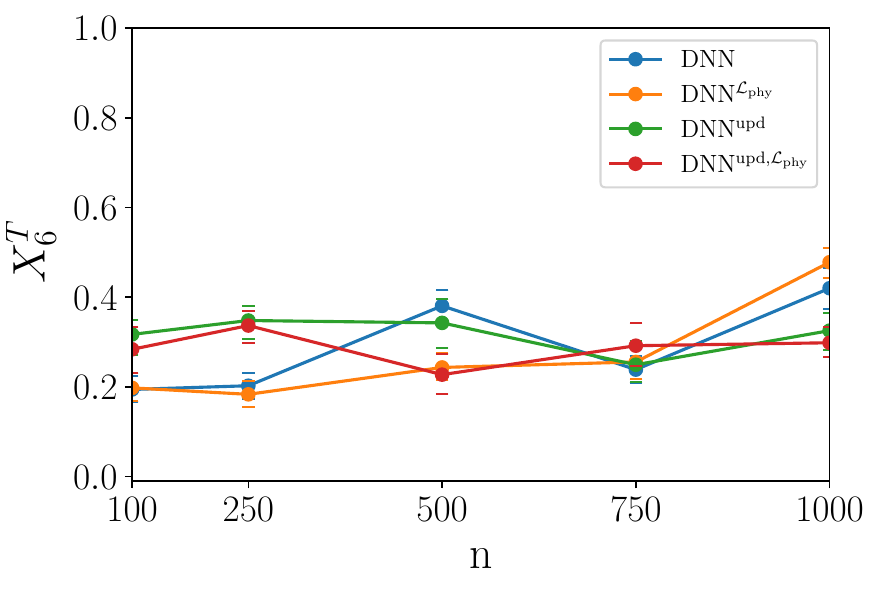} \\
    \includegraphics[width=.33\textwidth]{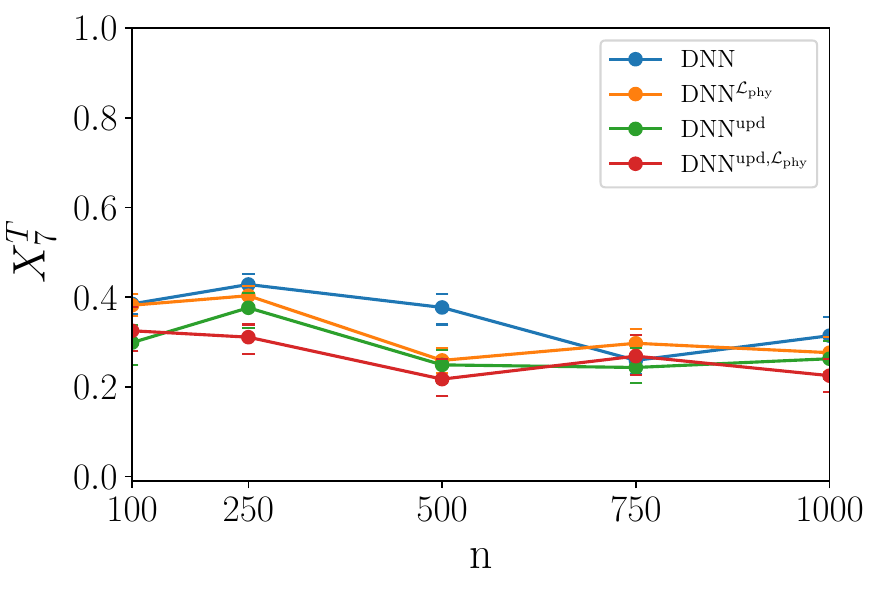} &
    \includegraphics[width=.33\textwidth]{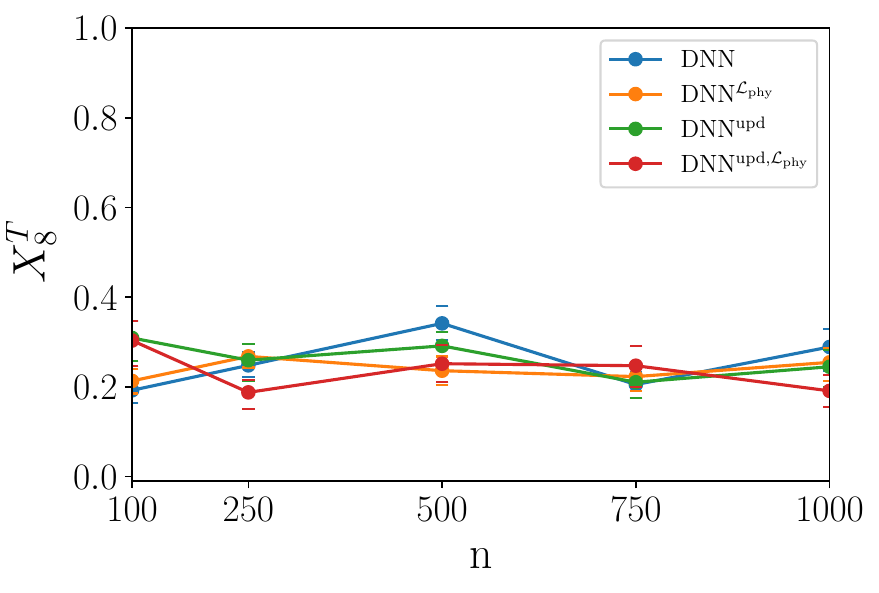} &
    \includegraphics[width=.33\textwidth]{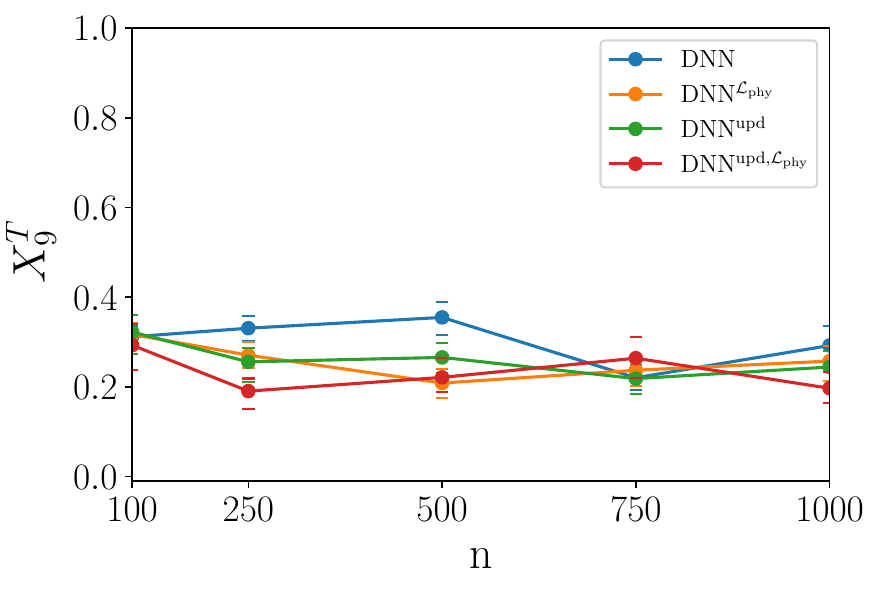} \\
    \includegraphics[width=.33\textwidth]{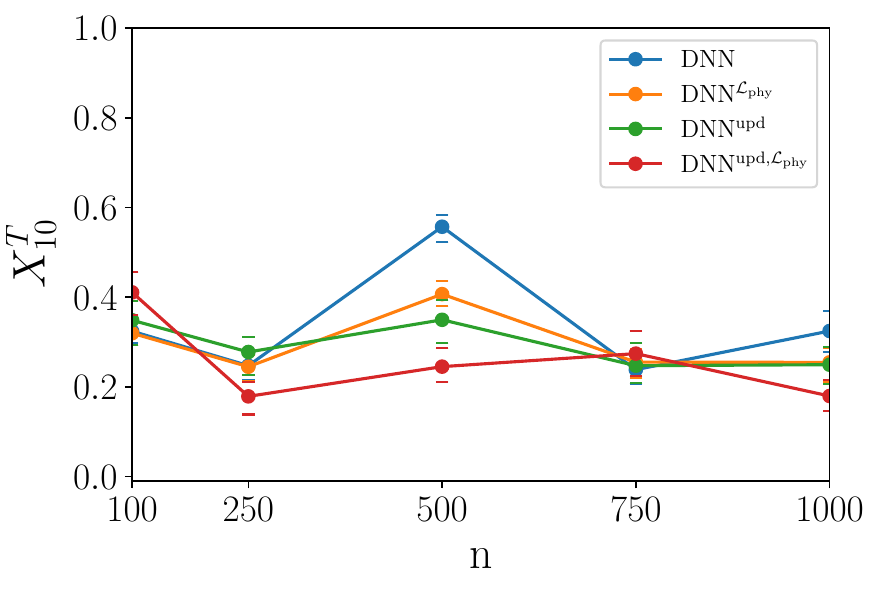} &
    \includegraphics[width=.33\textwidth]{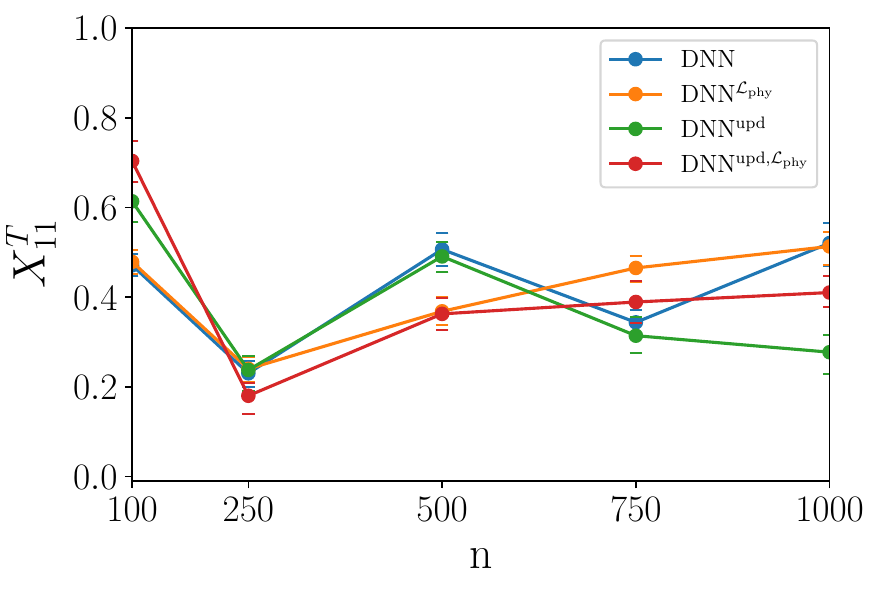} 
  \end{tabular}
  \caption{Total effect sensitivity index estimates for the eleven variables of the lake temperature modeling using the DNN models.}
  \label{fig:DNN_lake_tot}
\end{figure}

The Sobol' index computations with the DNN models (5-8) are based on 5000 MC samples and 100 stochastic forward passes through the networks. The calculated first-order sensitivity estimates of eleven input variables, for different numbers of experimental observations (training data) $n=(100,250,500,750,1000)$ are shown in Figs.~\ref{fig:DNN_lake_first} and~\ref{fig:DNN_lake_tot}.

All DNN models converge to similar first-order and total effect sensitivity estimates for all eleven inputs, and these values are consistent with the results obtained using GP models. The total effect sensitivity estimates obtained using the basic GP and DNN models for most of the input variables converge to a slightly different value than the ones obtained using the proposed PIML models. Note that the mean estimates of the PIML models converge more smoothly than the basic ML models. 

The results show that prediction intervals obtained using all models show a similar trend. The 95\% bounds for all DNN models are significantly narrower than the ones obtained using the GP models. The results indicate that the mean estimates and prediction intervals obtained using the PIML models 4 and 8 ($\mathbf{\rm GP^{\rm upd, \mathcal{L}_{\rm phy}}}$ and $\mathbf{\rm DNN^{\rm upd, \mathcal{L}_{\rm phy}}}$) converge to the mean estimates and bounds obtained using 1000 number of observations for training the models faster than the other models. Note that the mean estimates for Models 4 and 8 ($\mathbf{\rm GP^{\rm upd, \mathcal{L}_{\rm phy}}}$ and $\mathbf{\rm DNN^{\rm upd, \mathcal{L}_{\rm phy}}}$) converge to their final values within 250 experimental observations for almost of the variables. The prediction bounds of DNN models do not change significantly as more experimental observations are used to train the models. However, the prediction bounds get smaller as more number of training data is used to train the GP models. Moreover, the prediction bounds of physics-informed models are slightly narrower than the basic GP model. For further clarity regarding uncertainty, numerical values of the prediction bounds of first-order sensitivity estimates of $X_1$ (day of year) obtained using 100 realizations of the GP models and 100 forward passes through the DNN models are shown in Table~\ref{table:Bounds_ex2}. 
\begin{table*}[!htb]
\small
\setlength\tabcolsep{1.2pt}
    \caption{First-order sensitivity estimate prediction bounds of $X_1$ (day of year) for different amounts of experimental training data, using the GP and DNN models.}
    \label{table:Bounds_ex2}
\centering{%
    \begin{tabular*}{\linewidth}{@{\extracolsep{\fill}}l|*{5}c@{\extracolsep{\fill}} | *{5}c@{\extracolsep{\fill}}}
    \hline
            Models & \multicolumn{5}{c|}{Lower 95\% confidence limit} & \multicolumn{5}{c}{Upper 95\% confidence limit} \\
             & n=100 & n=250 & n=500 & n=750 & n=1000 & n=100 & n=250 & n=500 & n=750 & n=1000 \\
    \hline
            1. $\mathbf{\rm GP}$ & 0.00 & 0.27 & 0.67 & 0.61 & 0.24 & 0.38 & 0.71 & 0.82 & 0.77 & 0.53 \\
            2. $\mathbf{\rm GP^{\mathcal{L}_{\rm phy}}}$ & 0.01 & 0.51 & 0.60 & 0.63 & 0.32 & 0.36 & 0.76 & 0.77 & 0.77 & 0.55 \\
            3. $\mathbf{\rm GP^{\rm upd}}$ & 0.54 & 0.48 & 0.70 & 0.65 & 0.49 & 0.82 & 0.78 & 0.82 & 0.80 & 0.68 \\
            4. $\mathbf{\rm GP^{\rm upd, \mathcal{L}_{\rm phy}}}$ & 0.46 & 0.40 & 0.59 & 0.55 & 0.42 & 0.69 & 0.66 & 0.69 & 0.68 & 0.57 \\
            5. $\mathbf{\rm DNN}$ & 0.06 & 0.14 & 0.08 & 0.45 & 0.16 & 0.08 & 0.16 & 0.12 & 0.49 & 0.19 \\
            6. $\mathbf{\rm DNN^{\mathcal{L}_{\rm phy}}}$ & 0.06 & 0.12 & 0.18 & 0.30 & 0.30 & 0.09 & 0.15 & 0.21 & 0.33 & 0.33 \\
            7. $\mathbf{\rm DNN^{\rm upd}}$ & 0.03 & 0.29 & 0.15 & 0.41 & 0.37 & 0.06 & 0.33 & 0.20 & 0.45 & 0.41 \\
            8. $\mathbf{\rm DNN^{\rm upd, \mathcal{L}_{\rm phy}}}$ & 0.00 & 0.30 & 0.19 & 0.29 & 0.32 & 0.02 & 0.33 & 0.23 & 0.34 & 0.37 \\
    \hline
    \end{tabular*}}%
\end{table*}

The results of these numerical examples could be summarized as follows:
\begin{itemize}
    \item The GP models required more computational effort than the DNN models, both in training and prediction.
    \item The DNN models were able to achieve higher accuracy and lower uncertainty in prediction, due to the optimization of dropout rate and number of training epochs. 
    \item The physics-informed ML models are able to achieve higher prediction accuracy than the basic ML models, especially when the amount of available experimental data is small.  
\end{itemize}

Overall, for these numerical examples, the DNN models gave higher accuracy and lower uncertainty in the GSA results than the GP models, and also required less computational effort both in training and prediction. However, the first numerical example consisted of only two inputs and a single output. As the number of inputs and outputs increase with the second numerical example, all the ML models considered above were challenged w.r.t. adequacy of training data, computational effort in training and prediction, and the accuracy and uncertainty of the sensitivity estimates. The results showed that the mean estimates of the PIML Models 4 and 8 ($\mathbf{\rm GP^{\rm upd, \mathcal{L}_{\rm phy}}}$ and $\mathbf{\rm DNN^{\rm upd, \mathcal{L}_{\rm phy}}}$) for both numerical examples converge to the mean estimates obtained using the largest number of observations (i.e., 39 and 1000 observations for the first and second numerical example, respectively) for training the models faster than the other models.

\section{Conclusion}\label{Sec:Conclusion}
This paper developed methodologies for information fusion and machine learning for sensitivity analysis using both physics knowledge and experimental data, while accounting for model uncertainty. Variance-based sensitivity analysis is used to quantify the relative contribution of each uncertainty source to the variability of the output quantity. Two types of ML models were considered, namely, GP and DNN models. Several PIML models were developed by leveraging two strategies for incorporating physics knowledge into ML models: (1) incorporating physics constraints within the loss functions used in training the ML models, and (2) pre-training an ML model with simulation data and then updating it with experimental data. The first strategy does not use the physics model, whereas the second strategy does. 

The calculation of the Sobol’ indices with the GP model simply uses the proposed estimator (Eq.~\ref{eq:sens_est_gp}). On the other hand, with respect to the DNN model, we use the Monte Carlo dropout strategy to compute prediction bounds on the Sobol’ indices; previous work in this regard has only considered prediction bounds of the model output. Prediction bounds are computed for the sensitivity index estimates to account for the model uncertainty in the trained models, and the accuracy and computational effort of the various PIML models are compared.

The results show that the application of PIML strategies to both GP and DNN enables accurate Sobol’ index computations even with smaller amounts of experimental data while producing physically meaningful results. Thus, the proposed approach helps to fill the physics knowledge gap in the ML models while estimating the Sobol’ indices, by correcting for the approximation in the physics-based models. The numerical examples show that training the GP models and estimating the Sobol’ indices require more computational effort than the DNN models. The uncertainty regarding the sensitivity estimates obtained using the DNN models is smaller than the results obtained using the GP models. In the numerical examples, the DNN models are found to be more accurate compared to the GP models. The higher accuracy, lower uncertainty, and lower computational effort of the DNN models is attributed to their flexibility in terms of number of parameters, training epochs and dropout rate.

In future work, the proposed PIML approaches need to be tested for problems with a larger number of dimensions both in the input and output, with multiple combinations to further analyze the convergence of Sobol' index estimates. Future work can also explore the weighting of the two sources of information, since the data produced by physics-based models and experiments have different levels of credibility. The proposed approach for GP models can also be extended by using different kernels with varying smoothness. We note that only one kernel, namely the Gaussian kernel, is used in this paper. In order to have a more complete comparison between the GP and DNN models, different kernels with varying smoothness can be considered for future work.

\bibliographystyle{unsrt}  
\bibliography{references}

\end{document}